\documentclass[twocolumn,amsmath,trackchanges]{aastex702}

\usepackage{graphicx}%
\usepackage{multirow}%
\usepackage{amsmath,amssymb,amsfonts}%
\usepackage{amsthm}%
\usepackage{mathrsfs}%
\usepackage[title]{appendix}%
\usepackage{xcolor}%
\usepackage{textcomp}%
\usepackage{booktabs}%
\usepackage{algorithm}%
\usepackage{algorithmicx}%
\usepackage{algpseudocode}%
\usepackage{listings}%

\usepackage{acro}
\usepackage{hyperref}
\hypersetup{
    colorlinks,
    linkcolor={color1},
    citecolor={color1},
    urlcolor={color1}
}
\usepackage{cleveref}
\usepackage{xspace}

\acsetup{
  single    = false,
  first-style = short-long,
  list/sort  = true,
  cite/group = true,
  cite/group/cmd = \citealt,
  cite/group/pre = {; \xspace},
  patch/longtable=false,
}

\newcommand{\amigo}{\textsc{Amigo}\xspace}
\newcommand{\drpangloss}{\textsc{drpangloss}\xspace}

\newcommand\dlux{\textsc{$\partial$Lux}\xspace}
\newcommand\jax{\textsc{Jax}\xspace}

\definecolor{color1}{HTML}{1f77b4}
\definecolor{color2}{HTML}{ff7f0e}
\definecolor{color3}{HTML}{2ca02c}
\definecolor{color4}{HTML}{d62728}

\DeclareAcronym{amigo}{
  short = \amigo,
  long = Aperture Masking Interferometry Generative Observations,
  cite = {desdoigtsAmigoDatadrivenCalibration2026,charlesImageReconstructionJWST2026c}
}

\DeclareAcronym{psf}{
  short = PSF,
  long = Point Spread Function
}
\DeclareAcronym{otf}{
  short = OTF,
  long = Optical Transfer Function
}
\DeclareAcronym{ifu}{
  short = IFU,
  long = Integral Field Unit
}
\DeclareAcronym{fft}{
  short = FFT,
  long = Fast Fourier Transform
}
\DeclareAcronym{rom}{
  short = ROM,
  long = Reduced Order Modeling,
  cite = {Benner2015}
}
\DeclareAcronym{autodiff}{
  short = autodiff,
  long = Automatic Differentiation,
  cite={autodiff}
}
\DeclareAcronym{jwst}{
  short = JWST,
  long = the James Webb Space Telescope
}

\DeclareAcronym{niriss}{
  short = NIRISS,
  long = the Near Infrared Imager and Slitless Spectrograph,
  cite = {niriss1,niriss2}
}

\DeclareAcronym{nir}{
  short = NIR,
  long = Near Infrared
}

\DeclareAcronym{ami}{
  short = AMI,
  long = the Aperture Masking Interferometer,
  cite = {sivaramakrishnanPlanetarySystemStar2009a,sivaramakrishnanInfraredImagerSlitless2023},
}

\DeclareAcronym{bfe}{
  short = BFE,
  long = Brighter-Fatter Effect,
  cite = {Hirata2020,Choi2020}
}

\DeclareAcronym{edm}{
  short = EDM,
  long = Effective Detector Model
}
\DeclareAcronym{ddt}{
  short = DDT,
  long = Director's Discretionary Time
}

\DeclareAcronym{hpc}{
  short = HPC,
  long = High Performance Computing
}

\DeclareAcronym{nn}{
  short = NN,
  long = Neural Network
}
\DeclareAcronym{cnn}{
  short = CNN,
  long = Convolutional Neural Network
}
\DeclareAcronym{ml}{
  short = ML,
  long = Machine Learning
}
\DeclareAcronym{fov}{
  short = FOV,
  long = field of view
}

\DeclareAcronym{h2rg}{
  short = H2RG,
  long = Teledyne HAWAII-2RG
}
\DeclareAcronym{adc}{
  short = ADC,
  long = Analogue to Digital Converter
}
\DeclareAcronym{eso}{
  short = ESO,
  long = European Southern Observatory
}
\DeclareAcronym{vlti}{
  short = VLTI,
  long = Very Large Telescope Interferometer
}

\DeclareAcronym{ipc}{
  short = IPC,
  long = Inter-Pixel Capacitance
}
\DeclareAcronym{disco}{
  short = DISCO,
  long = Delay-Insensitive Subspace of Calibrated Observables
}
\DeclareAcronym{nrm}{
  short = NRM,
  long = Non-Redundant Mask
}
\DeclareAcronym{hst}{
  short = HST,
  long = Hubble Space Telescope
}
\DeclareAcronym{mcmc}{
  short = MCMC,
  long = Monte Carlo Markov Chain,
  cite={Metropolis1953}
}
\DeclareAcronym{iwa}{
  short = IWA,
  long = Inner Working Angle
}
\DeclareAcronym{opd}{
  short = OPD,
  long = Optical Path Difference
}
\DeclareAcronym{go}{
  short = GO,
  long = General Observer
}
\DeclareAcronym{sgd}{
  short = SGD,
  long = stochastic gradient descent,
  cite= {Ruder2016}
}
\DeclareAcronym{hmc}{
  short = HMC,
  long = Hamiltonian Monte Carlo,
  cite= {Betancourt2017}
}
\DeclareAcronym{mast}{
  short = MAST,
  long = the Mikulski Archive for Space Telescopes
}

\DeclareAcronym{gpu}{
  short = GPU,
  long = Graphics Processing Unit
}

\DeclareAcronym{au}{
  short = au,
  long = Astronomical Unit
}

\newcommand{\N}{\ensuremath{\mathcal{N}}}
\newcommand{\TN}{\ensuremath{\mathcal{TN}}}
\newcommand{\U}{\ensuremath{\mathcal{U}}}

\begin{document}

\title[Article Title]{A Candidate Innermost Fifth Planet In the HR~8799 System Revealed By JWST NIRISS Aperture Masking Interferometry}

\author[0000-0002-9242-9052, gname=Jayke,sname=Nguyen]{Jayke S. Nguyen}
\affiliation{Department of Astronomy \& Astrophysics, University of California, San Diego, La Jolla, CA 92093, USA}
\email{jsn001@ucsd.edu}

\author[0000-0002-1015-9029, gname=Louis,sname=Desdoigts]{Louis Desdoigts}
\affiliation{Leiden Observatory, Niels Bohrweg 2, Leiden, 2300RA, The~Netherlands}
\email{desdoigts@strw.leidenuniv.nl}

\author[0000-0002-7162-8036, gname=Alexandra,sname=Greenbaum]{Alexandra Z. Greenbaum}
\affiliation{IPAC, Caltech, 1200 East California Boulevard, Pasadena, CA 91125, USA}
\email{azg@ipac.caltech.edu}

\author[0000-0003-2595-9114, gname=Benjamin,sname=Pope]{Benjamin J.~S. Pope}
\affiliation{Astronomy and Space Technology Research Centre, School of Mathematical and Physical Sciences, Astrophysics and Space Technologies Research Centre, Macquarie University, Sydney, NSW, 2109, Australia}
\email{benjamin.pope@mq.edu.au}

\author[0009-0003-5950-4828, gname=Max,sname=Charles]{Max Charles}
\affiliation{Sydney Institute for Astronomy, School of Physics, University of Sydney, Camperdown, NSW 2006, Australia}
\email{max.charles@sydney.edu.au}

\author[0000-0002-9936-6285, gname=Quinn,sname=Konopacky]{Quinn M. Konopacky}
\affiliation{Department of Astronomy \& Astrophysics, University of California, San Diego, La Jolla, CA 92093, USA}
\email{qkonopacky@ucsd.edu}

\author[0009-0009-2223-2404, gname={Kaitlyn},sname={Hessel}]{Kaitlyn Hessel}
\affiliation{Department of Physics and Astronomy, University of Victoria, 3800 Finnerty Road, Elliot Building, Victoria, BC V8P 5C2, Canada}
\email{khessel@uvic.ca}

\author[0000-0001-9582-4261, gname={Dori},sname={Blakely}]{Dori Blakely}
\affiliation{Department of Physics and Astronomy, University of Victoria, 3800 Finnerty Road, Elliot Building, Victoria, BC V8P 5C2, Canada}
\affiliation{NRC Herzberg Astronomy and Astrophysics, 5071 West Saanich Road, Victoria, BC V9E 2E7, Canada}
\email{blakelyd@uvic.ca}

\author[0000-0003-1863-4960, gname={Matthew},sname={De Furio}]{Matthew De Furio}
\affiliation{Department of Astronomy, The University of Texas at Austin, 2515 Speedway, Stop C1400, Austin, TX 78712, USA}
\email{defurio@utexas.edu}

\author[0000-0001-5173-2947, gname={Clarissa},sname={Do \'{O}}]{Clarissa R. Do \'{O}}
\affiliation{Department of Astronomy, California Institute of Technology, Pasadena, CA 91125, USA}
\email{cdoo@caltech.edu}

\author[0000-0001-5485-4675, gname={Ren\'e},sname={Doyon}]{Ren\'e Doyon}
\affiliation{Trottier Institute for Research on Exoplanets, D\'epartement de Physique, Universit\'e de Montr\'eal, 1375 Avenue Th\'er\`ese-Lavoie-Roux, Montr\'eal, QC H2V 0B3, Canada}
\affiliation{Observatoire du Mont-M\'egantic, Universit\'e de Montr\'eal, Montr\'eal H3C 3J7, Canada}
\email{rene.doyon@umontreal.ca}

\author[0000-0002-6773-459X, gname={Doug},sname={Johnstone}]{Doug Johnstone}
\affiliation{Department of Physics and Astronomy, University of Victoria, 3800 Finnerty Road, Elliot Building, Victoria, BC V8P 5C2, Canada}
\affiliation{NRC Herzberg Astronomy and Astrophysics, 5071 West Saanich Road, Victoria, BC V9E 2E7, Canada}
\email{doug.johnstone@nrc-cnrc.gc.ca}

\author[0000-0003-2769-0438, gname={Jens},sname={Kammerer}]{Jens Kammerer}
\affiliation{European Southern Observatory, Karl-Schwarzschild-Straße 2, 85748 Garching, Germany}
\email{jkammere@eso.org}

\author[0000-0002-6780-4252, gname={David},sname={Lafreni\`ere}]{David Lafreni\`ere}
\affiliation{Trottier Institute for Research on Exoplanets, D\'epartement de Physique, Universit\'e de Montr\'eal, 1375 Avenue Th\'er\`ese-Lavoie-Roux, Montr\'eal, QC H2V 0B3, Canada}
\email{david.lafreniere@umontreal.ca}

\author[0000-0003-1212-7538,gname={Bruce},sname={Macintosh}]{Bruce A. Macintosh}
\affiliation{Department of Astronomy and Astrophysics, University of California, Santa Cruz, Santa Cruz, CA 95064, USA}
\affiliation{University of California Observatories, 1156 High Street, Santa Cruz, CA 95064, USA}
\email{bamacint@ucsc.edu}

\author[0000-0003-1227-3084, gname={Michael},sname={Meyer}]{Michael R. Meyer}
\affiliation{Astronomy Department, University of Michigan, Ann Arbor, MI 48109, USA}
\email{mrmeyer@umich.edu}

\author[0000-0001-6975-9056,gname={Eric},sname={Nielsen}]{Eric L. Nielsen}
\affiliation{Department of Astronomy, New Mexico State University, P.O. Box 30001, MSC 4500, Las Cruces, NM 88003, USA}
\email{nielsen@nmsu.edu}

\author[0000-0003-2461-6881,gname={Anne},sname={Peck}]{Anne E. Peck}
\affiliation{Department of Astronomy, New Mexico State University, P.O. Box 30001, MSC 4500, Las Cruces, NM 88003, USA}
\email{annepeck@nmsu.edu}

\author[0009-0008-9687-1877,gname={William},sname={Roberson}]{William Roberson}
\affiliation{Department of Astronomy, New Mexico State University, P.O. Box 30001, MSC 4500, Las Cruces, NM 88003, USA}
\email{wcroberson2000@gmail.com}

\author[0000-0003-1251-4124, gname={Anand},sname={Sivaramakrishnan}]{Anand Sivaramakrishnan}
\affiliation{Space Telescope Science Institute, 3700 San Martin Drive, Baltimore, MD 21218, USA}
\affiliation{Astrophysics Department, American Museum of Natural History, 79th Street at Central Park West, New York, NY 10024, USA}
\affiliation{Department of Physics and Astronomy, Johns Hopkins University, 3701 San Martin Drive, Baltimore, MD 21218, USA}
\email{anand@stsci.edu}

\author[0000-0001-7026-6291,gname={Peter},sname={Tuthill}]{Peter Tuthill}
\affiliation{Sydney Institute for Astronomy, School of Physics, University of Sydney, Camperdown, NSW 2006, Australia}
\email{peter.tuthill@sydney.edu.au}

\author[0000-0002-5922-8267, gname={Thomas},sname={Vandal}]{Thomas Vandal}
\affiliation{Trottier Institute for Research on Exoplanets, D\'epartement de Physique, Universit\'e de Montr\'eal, 1375 Avenue Th\'er\`ese-Lavoie-Roux, Montr\'eal, QC H2V 0B3, Canada}
\email{thomas.vandal@umontreal.ca}

\author[0000-0001-7591-2731, gname={Marie},sname={Ygouf}]{Marie Ygouf}
\affiliation{Jet Propulsion Laboratory, California Institute of Technology, 4800 Oak Grove Dr., Pasadena, CA 91109, USA}
\email{marie.ygouf@jpl.nasa.gov}


\begin{abstract}
We detect a candidate fifth planet in the HR~8799 system, directly imaged with the JWST/NIRISS Aperture Masking Interferometer (AMI). The detected source lies just above a $3\,\sigma$ contrast curve at a contrast of $\sim2\times10^{-4}$ in the F380M filter at a projected separation of $\sim150$ mas, corresponding to a few-to-several Jupiter mass planet at an orbital radius of $\sim 7$\,au. The separation of the candidate is compatible with absolute proper-motion constraints from Gaia and Hipparcos assuming it is bound, while its orbital position lies near a stable orbital solution of a fifth planet in a 3:1 mean motion resonance with planet e. This detection was made possible by a new JWST/NIRISS AMI data pipeline that reaches the photon noise limited potential of AMI by accounting for the optical and electronic systematics that limited sensitivity in prior analyses. Confirmation of this candidate would make HR~8799 the first directly imaged five-planet system and provide insight into the orbital dynamics and the dynamical evolution of planetary systems with widely-separated gas giants.
\end{abstract}


\section{Introduction} \label{sec1}

HR~8799 is one of the most unique and intriguing multi-planet systems discovered to date, and is the only directly-imaged system known to harbor 4 giant ($5-10$\,$M_\text{Jup}$) exoplanets orbiting their host star \citep{maroisDirectImagingMultiple2008c, maroisImagesFourthPlanet2010a}. The system is relatively young, with an estimated age of $\approx$ 40 Myr \citep{faramazDetailedCharacterizationHR2021}, with the four known planets orbiting at distances of $15-70$\,au. These young planets have long been the cornerstone of direct imaging, as their existence has challenged our understanding of planet formation. At first glance, the stability of a four-planet system with planet masses heavier than Jupiter seems improbable, as N-body dynamics are highly likely to eject planets in the system over long timescales. However, these planets are likely to be in mean motion resonance (MMR) in a Laplace 1:2:4:8 period ratio chain \citep{fabryckySTABILITYDIRECTLYIMAGED2010, gotbergLongtermStabilityHR2016a, gozdziewskiExactGeneralizedLaplace2020a}. MMR enables the survival of the system, creating islands of stability at integer period ratios between planets. Indeed as we reach longer time baselines from the discovery epoch, relative astrometry confirms that the system is likely in resonance, with the some of the most recent results given by \cite{zurloOrbitalDynamicalAnalysis2022}.

The system has long been speculated to host a fifth inner exoplanet \citep{skemerFIRSTLIGHTLBT2012, maireLEECHExoplanetImaging2015, wahhajSearchFifthPlanet2021, zurloOrbitalDynamicalAnalysis2022, thompsonDeepOrbitalSearch2022}, echoing the discovery of the fourth companion, HR~8799~e \citep{maroisImagesFourthPlanet2010a}, which substantially reshaped our understanding of the system's architecture. A fifth planet may continue the system's MMR chain \citep{gozdziewskiMultipleMeanMotion2014a}, occupying an island of stability in a 3:1 or 2:1 resonance relative to the fourth innermost planet. These resonant stable orbital solutions posit a $7-9\,M_\text{Jup}$ innermost planet and are stable over $\sim 1$\,Gyr timescales.

Our imaging probes have not been able to reach this posited fifth planet due to the high contrast at high angular resolution required to detect a few-to-several Jupiter mass companion at the system's innermost separation. Observations from the ground \citep[e.g.][]{skemerFIRSTLIGHTLBT2012, wahhajSearchFifthPlanet2021, thompsonDeepOrbitalSearch2022} have placed stringent constraints on an innermost planet, but have not established a confirmed fifth planet. A fifth planet is challenging to find because the putative source is near the inner working angle of those observations. Additionally, non-redundant masking (NRM) systems from the ground are not capable of reaching the contrasts required to see an inner planet, typically reaching limiting contrasts of $>10^{-3}$ \citep{Greenbaum2019}. The presence of an additional inner planet remains uncertain, motivating complementary searches with \ac{niriss} \ac{ami} to provide additional constraints on the inner system.

The JWST/NIRISS \ac{ami} was designed with science cases like this in mind, achieving interferometric resolution at high dynamic range, approaching a theoretical inner working angle of ${\lambda}/{2D}$. The strong recovery of PDS~70~b and~c with JWST \ac{ami} \citep{blakelyJamesWebbInterferometer2025b} demonstrates the power of this technique to probe angular scales inaccessible to conventional direct imaging. Given the potential sensitivity of \ac{ami}, an early JWST Guaranteed Time Observation program focused on finding faint companions ($\sim10^{-4}$ contrast) at short separations ($<{\lambda}/D$) in systems with known companions, including HR~8799. Recent advances in \ac{ami} interferometric forward modeling allow us to revisit this data and test the longstanding prediction of a fifth innermost planet in the system.

We present a candidate inner fifth planet in the HR~8799 system, HR~8799~f, in public archival JWST \ac{ami} data. The object remained unseen for a significant amount of time in previous investigations due to prior limitations in data analysis. We use \ac{amigo}, a new \ac{ami} data processing pipeline to forward model the optical path and detector response, we achieve the contrast required to recover the candidate above a $3\,\sigma$ contrast curve in a single-band observation. The detection is only possible with the improved calibration and contrast performance enabled by \ac{amigo}'s forward-modeling approach. In Section~\ref{sec:obs} we outline the observations taken with JWST. In Section~\ref{sec:methods} we discuss previous limitations in data processing and the new data reduction process with \ac{amigo}. We discuss our findings in Section~\ref{sec:res} analyzing both astrometry with orbit fits and photometry with comparisons to previous works. Lastly, in Section~\ref{sec:disc} we highlight the implications of this candidate, discuss previous studies of the system, and suggest improvements to future observations of HR~8799 with JWST \ac{ami}.

\section{Observations}\label{sec:obs}

The observations were obtained from public archival JWST data of the HR~8799 system from the NIRISS instrument team's Guaranteed Time Observation program GTO~1200, available on \ac{mast} and may be found at: \dataset[10.17909/a016-yj89]{http://dx.doi.org/10.17909/a016-yj89}. The exposures were taken on 2023~August~3~(UT) using the \ac{ami} mode on \ac{niriss}, inserting the non-redundant mask in the instrument pupil plane, using \texttt{NISRAPID} readout mode across the \texttt{SUB80} subarray of the detector. The observations of HR~8799 have 1553.46 seconds of exposure time, 6864 integrations, over 3 groups covering only the F380M filter. A calibrator star HD 93649 was also observed in the same filter, using exposure settings to match a similar well depth as the science observations. The calibrator data had 2780.72 seconds of exposure time, 9215 integrations, and 4 groups. Notably, these observations only encompass a single dither, as opposed to the currently recommended subpixel dithering scheme \citep{desdoigtsAmigoDatadrivenCalibration2026, charlesImageReconstructionJWST2026c} that aids in separating detector effects from astrophysical sources. Reduction was performed entirely using \ac{amigo}, which uses raw, unprocessed \texttt{Stage 0} JWST data products.


\section{Methods}\label{sec:methods}

\ac{ami} is one of the highest angular resolution imaging modes on JWST, providing a 7-hole non-redundant aperture mask in the pupil wheel in the \ac{niriss} instrument. The non-redundant mask means both that phase retrieval, at least of the \ac{opd} piston terms on each sub-aperture, is a well-posed problem \citep{Cheetham2012}; and it provides self-calibrating closure phase observables \citep{closure_phase} constructed as linear combinations of Fourier-plane measurements, which are invariant under piston perturbation by \acp{opd} between the subapertures.

After commissioning \ac{ami}, it was found that these observables as extracted from previously-existing pipelines were limited by systematics from the \ac{bfe} in the detector \citep{Sallum2023,Ray2023b}. Its performance on bright targets was limited to contrasts $>10^{-3}$ \citep{sivaramakrishnanInfraredImagerSlitless2023}, an order of magnitude below the pre-flight performance predictions \citep{greenbaumIMAGEPLANEALGORITHMJWSTS2014}. The high spatial frequency fringes of the AMI point spread function (PSF) make this effect particularly pertinent, as bright pixels affect all of the neighboring pixels and distort the fringe pattern, compromising the recovered complex visibilities. This non-linear blur cannot be represented as a single convolution \citep{Hirata2020,bfe_miri}, and therefore does not calibrate in ordinary interferometric pipelines.

\ac{amigo} was designed as a qualitatively different pipeline to overcome these limitations \citep{desdoigtsAmigoDatadrivenCalibration2026,charlesImageReconstructionJWST2026c}. It is a complete end-to-end model of \ac{niriss} \ac{ami} on JWST, modeling the complete optical and electronics system from photons-to-pixels, extending to data post-processing. \ac{amigo} has allowed \ac{ami} to reach the pre-flight expected contrast limits on the order of $10^{-4}$ and has been used to detect and confirm HD~206893~c \citep{desdoigtsAmigoDatadrivenCalibration2026} at high contrast ($\sim2\times10^{-4}$) and small separation ($\sim$100\,mas) using multi-filter (F380M, F430M, F480M) observations.

The forward model it applies is highly parametrized and physically informed, including optical aberrations represented with a physical-optics propagation model in the differentiable framework \dlux \citep{dlux2}, and an ``effective detector model'' for the electronics to directly predict uncalibrated pixel reads up-the-ramp, using Bayesian inference to infer free parameters. The effective detector model applies local linear and nonlinear gains and a neural network that encodes the \ac{bfe} dynamics, and is held fixed after being calibrated on an ensemble of point-source calibrator and flat-field datasets. We discuss the effects of keeping the detector model fixed and residual calibration uncertainties later in Section~\ref{sec:consistency}.

The telescope \ac{opd} state is inferred directly from the data being analyzed, jointly with the astrophysical source properties. More specifically, the \ac{opd} is parameterized by 10 Zernike modes per subaperture, giving a total of 70 coefficients across all subapertures in the non-redundant mask. Astrophysical properties such as the source position, flux, and complex visibility are fit simultaneously in the same gradient-based optimization. Phase and amplitude aberrations originating from defocus and distortion in the non-redundant apertures are well-determined from the original \ac{amigo} calibration dataset and held fixed in the pipeline. Even with thousands of parameters, this optimization problem is efficient and well-posed thanks to its use of natural gradient descent \citep{Amari1998} and efficient derivatives implemented in \jax \citep{jax}.

\ac{amigo} can be operated as an interferometric pipeline or directly for image-plane deconvolution \citep{charlesImageReconstructionJWST2026c}, producing visibilities represented in a latent basis, using \ac{disco} self-calibration properties to generalize closure phase. \ac{disco} observables are self-calibrated combinations of visibility amplitudes and phases that suppress the first-order effects of optical aberrations while retaining astrophysical information. First we use Bayesian inference to extract best-fit complex visibilities and their covariance matrices. The Jacobian of the visibility observables with respect to wavefront calibration is used to null out vulnerable modes in the same manner as with kernel phases and amplitudes \citep{Martinache2010,Pope2016}. Uncertainties from both datasets are propagated directly into the calibrated observables, and then the covariance matrix is diagonalized to provide statistically-independent observables \citep[following][]{irelandPhaseErrorsDiffractionlimited2013} -- the \acp{disco}. For the science target and calibrator, we average the projected observables over exposures and subtract the calibrator contribution from the science measurements.


We use the differentiable interferometry modeling pipeline \href{https://github.com/benjaminpope/drpangloss}{\drpangloss} \citep{blakelyJamesWebbInterferometer2025b} to perform Bayesian inference about the astrometry and photometry of companions from these interferometric data assuming a constant planet-to-star flux ratio across the bandpass, evaluating the model visibilities at the filter's effective wavelength. Lastly, we use a single companion visibility model to perform a coarse grid search to locate faint companions, followed by a fine grid search using \ac{hmc} and multi-companion visibility models to precisely determine their properties.

\section{Results}\label{sec:res}

\begin{figure*}
\centering
\includegraphics[width=0.99\textwidth]{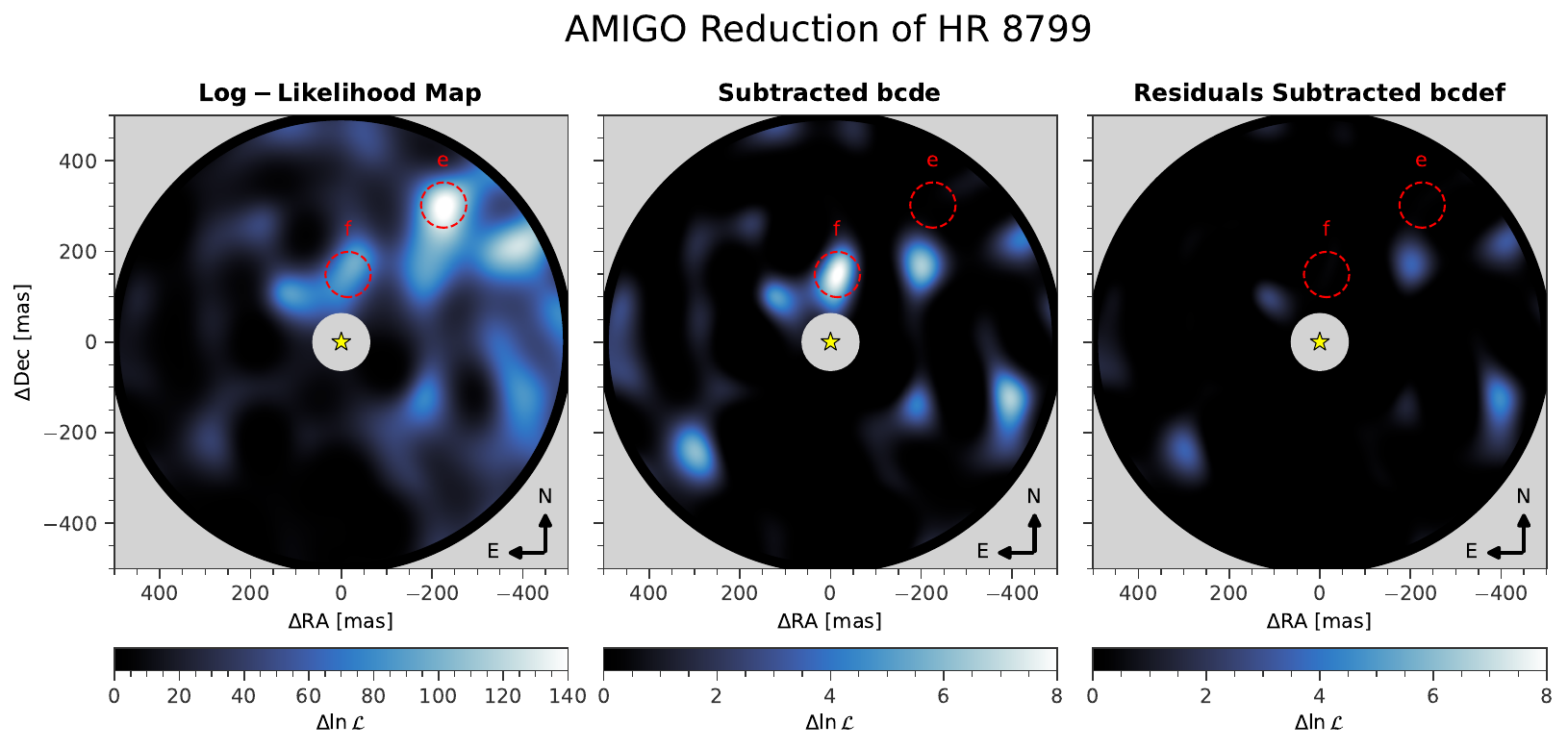}
\caption{Delta-log-likelihood maps for companion presence in the \ac{ami} F380M data of HR~8799, where we define $\Delta\ln\mathcal{L} = \ln\mathcal{L}_\text{Companion} - \ln\mathcal{L}_\text{Null}$, the log-likelihood difference between the presence of a companion and no companion (e.g. the null hypothesis). In the left panel we show fits directly to the DISCO data from \drpangloss using a single companion model, with a clear and statistically-significant maximum at the expected position of planet e, and a local maximum closer to the star at the position of the f candidate. In the middle we show a residual log-likelihood map obtained from fitting a single companion model after subtracting the best-fit joint model of planets bcde. The candidate~f is an unambiguous and significant peak. On the right we show the residual log-likelihood map subtracting the joint model of planets bcde and candidate~f from the data. The other planets are outside of the rendered field of view, but present in the field of view of the \ac{ami} data themselves. For visual clarity we overplot red circles (of arbitrary radius) encompassing the companion signals. The center gray region masks the interferometric inner working angle of $\lambda/2D$.}\label{fig:maps}
\end{figure*}

\begin{figure*}
\centering
\includegraphics[width=0.9\textwidth]{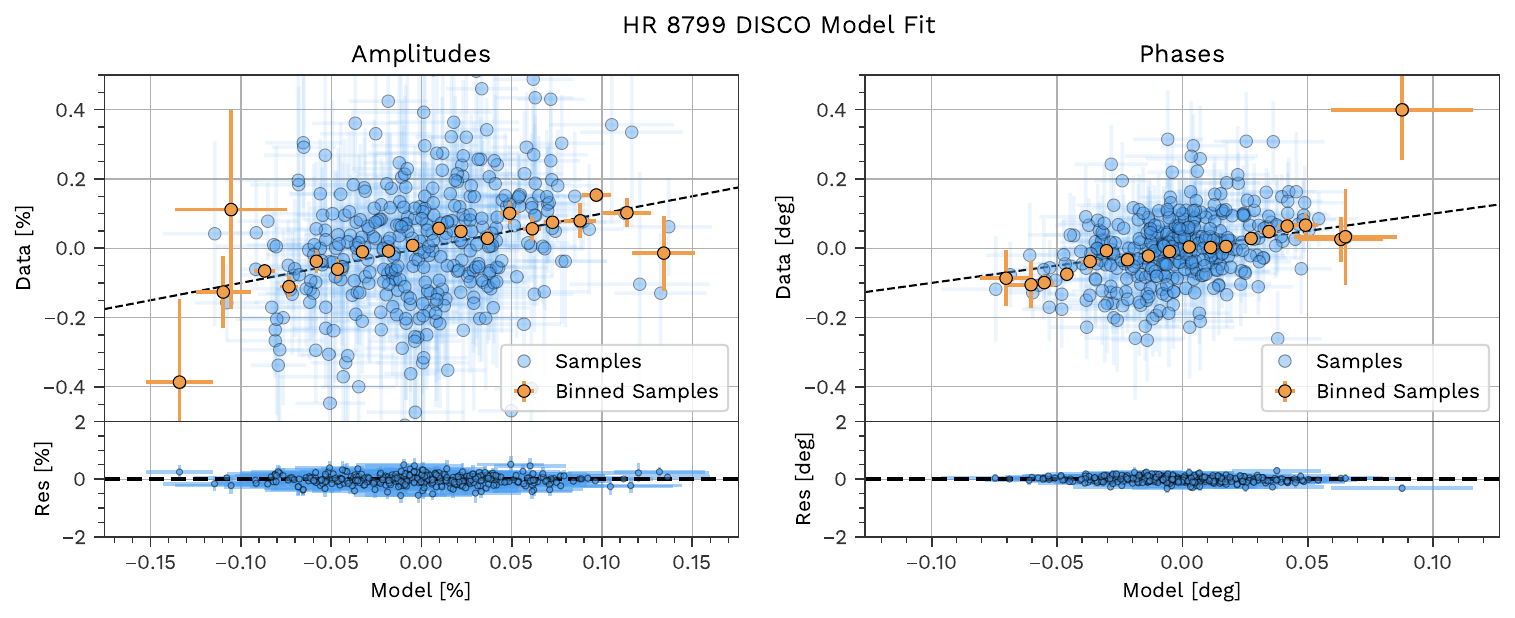}
\caption{Posterior-predictive correlation plots of our model of the complex fringe visibilities with the amplitude (left) and phase (right) in the DISCO basis. The left shows visibility amplitudes in units of percent, and the right shows phase components in degrees. The dashed line in the upper plots marks the one-to-one relationship between the data and model. These are very noisy individually and are binned for visual clarity and are not a statistical operation, only a modest trend can be seen. The non-zero trend in the binned values in both plots are consistent with the presence of faint companions. The error scaling is applied to each sample's uncertainties to ensure that $\chi_\nu^2 \approx 1$ in the final model fit. For our model, the error scaling in visibility amplitude is 2.233 and in phase is 1.857.}\label{fig:correlation_plots}
\end{figure*}

\begin{figure}
\centering
\includegraphics[width=0.49\textwidth]{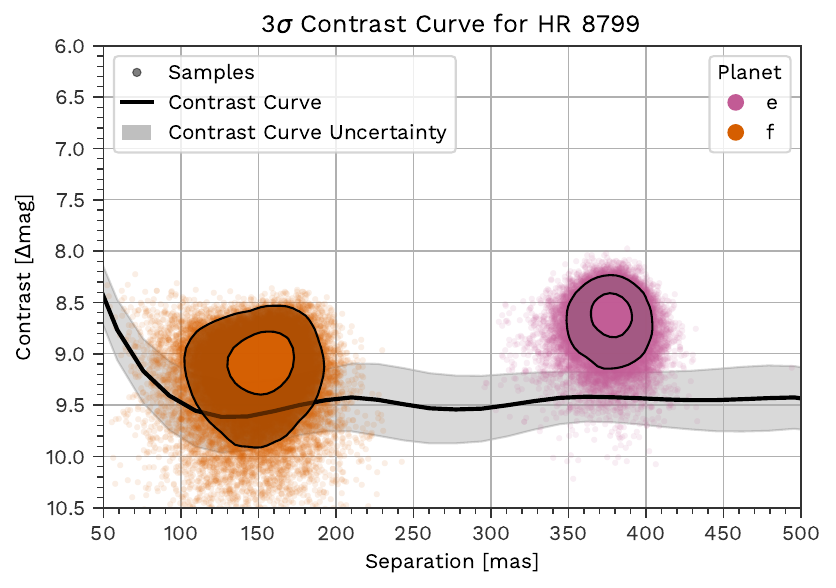}
\caption{Contrast curve for the HR~8799 \ac{ami} data using the \citet{ruffio_2018} method as implemented in \drpangloss, obtained from the residual after planets bcde and candidate~f were subtracted. The 68\% and 95\% posterior contours in contrast and separation are overlaid for planets~e and~f. The shaded region of the contrast curve represents the $\pm1\sigma$ uncertainty on the curve.}\label{fig:contrast_curve}
\end{figure}

To search for companions in the HR~8799 system, we use the \ac{amigo} pipeline to fit a single companion visibility model to our data in the \ac{disco} observable basis. We present the delta-log-likelihood maps of this fit to the innermost region of the system HR~8799 in Figure \ref{fig:maps} (left). We crop the innermost region here for clarity, but the field of view of \ac{niriss} \ac{ami} encompasses the entire system. As shown, we recover known planet e at a location near previous measurements and expected from orbital fits. Planets bcd are also present near expected locations outside the bounds of this figure.

We then jointly fit the four known planets in the \ac{disco} visibility basis. For each planet, we adopt uniform priors over $200$\,mas square centered over its predicted position at the AMI epoch. Predicted positions of the known planets were obtained using \texttt{whereistheplanet} \citep{wangWhereistheplanetPredictingPositions2021}. We use log-uniform priors in contrast between $10^{-5}$ and $10^{-2}$ and infer the planet parameters directly using \ac{hmc}. To search for additional companions, we subtract the joint four-planet model in \ac{disco} observable-space and refit the data with a single companion model. We obtain the delta-log-likelihood map of Figure \ref{fig:maps} (middle) revealing, a signal with $\Delta\ln\mathcal{L}\approx 8$ just north of the host star, which we interpret as a candidate innermost planet in the system, HR~8799 f. We then construct a joint five-planet model, centering the position prior for candidate~f at the coarse location found in the delta-log-likelihood map. We use \ac{hmc} to jointly infer the position and contrast of planets bcde and candidate~f.

\begin{deluxetable}{cccc}
\tablecaption{Recovered F380M filter astrometry and photometry from GTO~1200 for known companions and candidate planet in the HR~8799 system. The median values and approximate $1\,\sigma$ uncertainties are listed.} \label{tab:astrophot}
\tablehead{
    \colhead{Planet} &
    \colhead{$\Delta$RA (mas)} &
    \colhead{$\Delta$Dec (mas)} &
    \colhead{Contrast ($\Delta$\,mag)}
}
\startdata
\midrule
b & $\phantom{-}1645.2 \pm 32.9$ & $\phantom{-}535.7 \pm 28.7$ & $8.97 \pm 0.21$ \\
c & $\phantom{0}{-291.6} \pm 15.2$ & $\phantom{-}917.7 \pm 16.2$ & $9.57 \pm 0.36$ \\
d & $\phantom{0}{-634.4} \pm 14.0$ & $-321.1 \pm 18.0$ & $8.91 \pm 0.18$ \\
e & $\phantom{0}{-225.8} \pm 10.4$ & $\phantom{-}302.2 \pm 13.3$ & $8.66 \pm 0.23$ \\
f & $\phantom{00}{-15.0} \pm 19.3$ & $\phantom{-}149.0 \pm 22.8$ & $9.14 \pm 0.39$ \\
\enddata
\end{deluxetable}

\begin{deluxetable}{ccccc}
\tablecaption{Astrometric consistency of our recovered \ac{ami} values (from Table \ref{tab:astrophot}) in $\Delta\mathrm{RA}$ and $\Delta\mathrm{Dec}$ to median values from \texttt{whereistheplanet} (WITP) for the four known planets on 2023 August 3 (UT). Checks (\checkmark) indicate that the WITP value is within the specified consistency range (e.g. within 1, 2, or 3$\sigma$ error bars) of the \ac{ami} data and the $\times$ indicates that it is not within that range. Errors on positions from WITP are at most, $\sim2$\,mas and are omitted here.}\label{tab:witp_astrometry_comparison}
\tablehead{
    \colhead{} &
    \colhead{\textbf{WITP Data}} &
    \multicolumn{3}{c}{\textbf{Consistency}} \\[-1.0ex]
    \colhead{Planet} &
    \colhead{$(\Delta\mathrm{RA}, \Delta\mathrm{Dec})$ (mas)} &
    \colhead{$1\,\sigma$} &
    \colhead{$2\,\sigma$} &
    \colhead{$3\,\sigma$}
}
\startdata
\midrule
b & $(\phantom{-}1635.41,\, \phantom{-}532.42)$
  & $(\checkmark, \checkmark)$ & $(\checkmark, \checkmark)$ & $(\checkmark, \checkmark)$ \\
c & $(\phantom{0}{-288.59},\, \phantom{-}909.70)$
  & $(\checkmark, \checkmark)$ & $(\checkmark, \checkmark)$ & $(\checkmark, \checkmark)$ \\
d & $(\phantom{0}{-606.14},\, {-345.32})$
  & $(\times, \times)$ & $(\times, \checkmark)$ & $(\checkmark, \checkmark)$ \\
e & $(\phantom{0}{-231.65},\, \phantom{-}325.88)$
  & $(\checkmark, \times)$ & $(\checkmark, \times)$ & $(\checkmark, \checkmark)$ \\
\enddata
\end{deluxetable}

We present the median posterior value and approximate $1\,\sigma$ uncertainties for our measurements for all five objects in Table \ref{tab:astrophot}. These uncertainties are calculated as standard deviations of the corresponding posterior samples from the joint five-planet model. The marginal posterior distributions of the reported parameters are approximately Gaussian. The recovered photometry for the four known planets and planet candidate are given in terms of relative photometry ($\Delta$\,mag), since the forward modeling of \ac{amigo} is not accurately flux calibrated to an absolute source. We provide a comparison for our recovered astrometry to locations obtained from \texttt{whereistheplanet} in Table \ref{tab:witp_astrometry_comparison}, and find that planets b and c are consistent to within $1\,\sigma$ and d and e are consistent within $3\,\sigma$. The weaker agreement for planets d and e may reflect calibration errors in the detector model, which we will discuss later.


The reduction in likelihood after subtracting a joint five-planet model of the significant signals in the image as shown in Figure \ref{fig:maps} (right) indicate that there are unlikely to be more signals remaining within the field of view. In particular, the other maxima including the lobe to the left of the candidate and lobe between e and candidate~f are reduced to weak residual peaks ($\Delta \ln \mathcal{L}<4$) after subtracting the full model, suggesting that they are interferometric aliases or residuals in the image-plane rather than an actual source. Performing an \ac{hmc} fine grid search for a single companion model on these two maxima in the final residuals yield poorly constrained posteriors that reinforce our claim that these are not true astrophysical sources.

Following common interferometric model-checking practice, we perform a posterior-predictive check of the amplitude and phase of the complex fringe visibilities (as expressed in the \ac{disco} basis), for the full system (including planets bcde and candidate~f), shown in Figure \ref{fig:correlation_plots}. This check uses the full data before subtraction of the five-planet model. The figure shows a modest trend that supports the presence of faint companions in the data as we see in the delta-log-likelihood maps. If there were no companions present, the samples would scatter about zero and be consistent with a flat line.

We can compare the posteriors from the \ac{hmc} fit to the measured $3\,\sigma$ contrast curve calculated using a Bayesian framework \citep{ruffio_2018} and the residuals after removing the signals from planets bcde and candidate~f. The posteriors are overlaid for planet e and candidate~f in Figure \ref{fig:contrast_curve}. Planet e is clearly recovered, but candidate companion f lies only slightly above the $3\,\sigma$ contrast curve therefore only qualifying as a tentative, but statistically significant detection.

\subsection{Orbit Fits}

To determine the plausibility of the candidate fifth companion, we perform orbit fits of our recovered astrometry to deduce whether our recovered data may be consistent with astrometric measurements from Gaia and Hipparcos. The purpose of these fits are not to definitively measure the orbit, but to check whether our measured position is compatible with absolute astrometric measurements and to predict the future position of candidate~f for followup observations. We use \href{https://github.com/sefffal/Octofitter.jl}{\texttt{Octofitter}} \citep{Octofitter}, a Bayesian orbital fitting package using data from the \texttt{G23H} catalog \citep{G23H}, which incorporates Gaia and Hipparcos proper motion data and radial velocity measurements, using the Pigeons Non-Reversible Tempering sampler \citep{surjanovicPigeonsjlDistributedSampling2025}.

We combine astrometry from the following sources: absolute astrometry of the central star from Gaia and Hipparcos, relative astrometry on the positions of the planets from \cite{zurloOrbitalDynamicalAnalysis2022} (a compilation from multiple sources and instruments), an additional epoch from \cite{Balmer_2025}, a previously unpublished ground-based epoch from NIRC2 at the Keck Observatory on 2025 October 09, and lastly the GTO~1200 \ac{ami} epoch. In total, we fit 54, 54, 66, 52, and 1 astrometric data points for planets bcde and candidate~f, respectively, spanning in time from 1998 to 2025. Details on the 2025 Keck dataset may be found in the appendix. The two extra epochs beyond 2023 ensure that there is data after the JWST \ac{ami} observation to better constrain the fit. Additionally, we add a shared log-uniform jitter prior to each of our relative astrometric data points ranging from 1\,mas to 10\,mas. This jitter accounts for instrument-to-instrument differences, varying parallax measurements, and other systematics that could arise from the diversity of data reduction schemes used.

\subsubsection{Orbital Element Priors}

\begin{table*}
\centering
\caption{Underlying priors adopted for the stellar system-wide and planet orbital parameters in the \texttt{Octofitter} fit. Observable-based priors are later applied to the planet orbital element priors, modifying the priors we actually sample from during the fit.}
\label{tab:priors}
\small
\setlength{\tabcolsep}{4pt}
\begin{tabular}{cccccccc}
\hline\hline
\multicolumn{8}{c}{\textbf{System Parameters}} \\

Object
& $m$ ($M_{\odot}$)
& $i$ (deg)
& $\Omega$ (deg)
& $\mu_\alpha$ (mas/yr)
& $\mu_\delta$ (mas/yr)
& \multicolumn{2}{c}{$\pi$ (mas)} \\
\hline

\vspace{1.0em}

Star
& \TN(1.5, 1.0;\,0.75, 1.75)
& $\cos(i) = \mathcal{U}(-1,1)$
& \U(0, 360)
& \N(108, 10)
& \N(-50, 10)
& \multicolumn{2}{c}{\TN(24.462, 0.046;\,24.0, --)} \\

\multicolumn{8}{c}{\textbf{Planet Parameters}} \\

Planet
& $m$ ($M_{\text{Jup}}$)
& $e$
& $\omega$ (deg)
& $\theta$ (deg)
& Mutual $i$ (deg)
& $\Omega$ (deg)
& Period (yr) \\

\hline

b & \TN(4.585, 3.0;\,0.5, 20) & \U(0, 0.5) & \U(0, 360) & \U(0, 360) & \N(0, 5) & \U(0, 360) & \U(300, 800) \\
c & \TN(8.749, 3.0;\,0.5, 20) & \U(0, 0.5) & \U(0, 360) & \U(0, 360) & \N(0, 5) & \U(0, 360) & \U(120, 450) \\
d & \TN(7.818, 3.0;\,0.5, 20) & \U(0, 0.5) & \U(0, 360) & \U(0, 360) & \N(0, 5) & \U(0, 360) & \U(50, 220) \\
e & \TN(7.620, 3.0;\,0.5, 20) & \U(0, 0.5) & \U(0, 360) & \U(0, 360) & \N(0, 5) & \U(0, 360) & \U(20, 120) \\
f & \U(0.5, 20) & \U(0, 0.5) & \U(0, 360) & \U(0, 360) & \N(0, 5) & \U(0, 360) & \U(5, 50) \\

\hline
\end{tabular}
\par\smallskip
\begin{minipage}{\textwidth}
\raggedright
\footnotesize
\textit{Note.}
$\TN(\mu,\sigma;a,b)$ denotes a truncated normal distribution with mean $\mu$, standard deviation $\sigma$, and bounds $[a,b]$; $\N(\mu,\sigma)$ denotes a normal distribution; and $\U(a,b)$ denotes a uniform distribution over $[a,b]$. The system-level inclination and longitude of the ascending node define a shared reference plane and are jointly fit with the orbital parameters. All other orbital elements are fit independently within the model.
\end{minipage}
\end{table*}

To mitigate biases from incomplete orbital coverage, we apply observable-based priors \citep{oneilImprovingOrbitEstimates2019, dooOrbitalEccentricitiesDirectly2023a} to all of our orbital elements. The underlying priors on the stellar and planet properties can be found in Table \ref{tab:priors}. For the host star mass, the resulting orbit fit differs depending on whether we use stellar evolutionary models \citep{bainesCHARAARRAYANGULAR2012} or dynamical mass estimates \citep{sepulvedaDynamicalMassExoplanet2022}. Considering this, we use informed Gaussian priors on our host star mass, encompassing both measurements allowing the orbit fit to determine the host star mass. For independently accurately measured quantities such as the parallax and barycentric proper motion we use strong informative Gaussian priors. 

Ideally, the planet masses also use uninformative uniform priors to allow the data to determine the mass, but due to the large parameter space required to account for all of the orbital elements for each of the planets, the sampler typically struggles to converge. Additionally, since there is only one astrometric data point, the orbit for candidate f is very poorly constrained. An alternative approach would be to use atmosphere-retrieval-only estimations of the planetary mass, but these estimates are typically broad and vary significantly depending on the particular model used \citep{nasedkinFourofakindComprehensiveAtmospheric2024a}. While other studies pin the planet masses with atmospheric retrievals through constrained dynamical mass, these masses are determined under the presumption of a four-planet model.

To narrow the parameter space, we use the five-planet ``Va'' dynamically stable resonant model from \cite{gozdziewskiMultipleMeanMotion2014a} as our informative priors on the known planet masses. Out of all of the resonant five-planet models presented, the Va solution is the closest model to our recovered data. As we will discuss later, we do not consider the proximity of our measured positions to the predicted positions from the resonant solution as evidence for detection and consider it circumstantial. In our orbit fits, we adopt Gaussian mass priors for the known planets, centered on masses from this stable resonant solution. Each prior has a standard deviation of $\sigma = 3\,M_\text{Jup}$, representing nearby orbital solutions to the stable solution found. For candidate planet f, we use an uninformative uniform prior on the mass. The priors on the remaining orbital elements per-planet do not use the Va model solution. These priors instead use underlying uniform priors, with the exception of mutual incliation which uses a near-coplanarity Gaussian prior.


\subsubsection{Orbit Fit Results}

\begin{figure*}
\centering
\includegraphics[width=\textwidth]{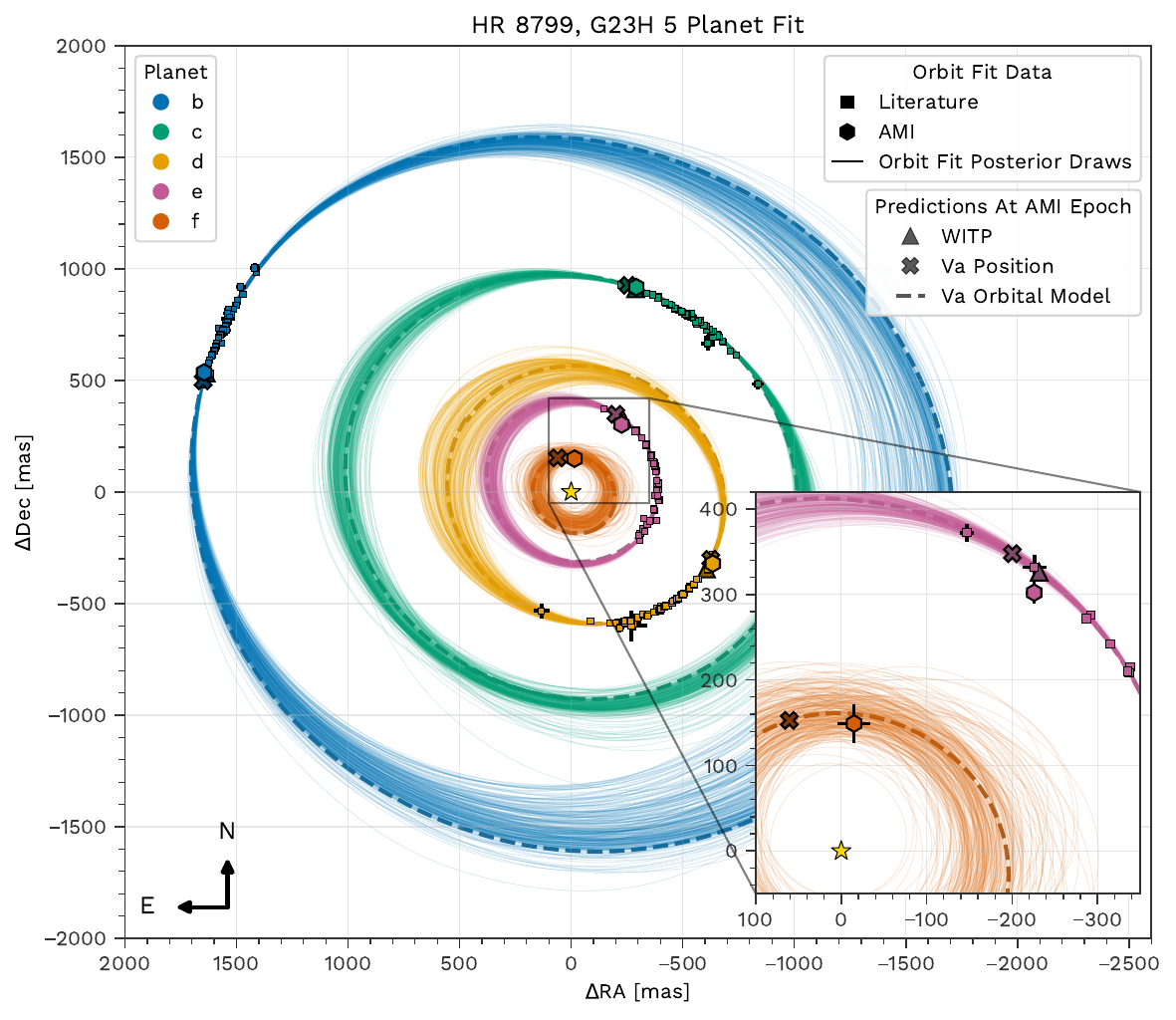}
\caption{Orbit fit results from \texttt{Octofitter} using all publicly available astrometry for HR~8799. Each planet is denoted in a different colour, with partly-transparent curves showing posterior orbit samples and filled markers show astrometry from literature and \ac{ami} data. The slightly grayer markers are predictions of planet positions at the \ac{ami} epoch from WITP and the resonant Va model, with the dashed curves showing the Va model orbit. Error bars are present on all data points and may be too small to see. The Va position and the WITP predictions are not used as data in our orbit fit and are plotted here for illustrative purposes.}\label{fig:hr_8799_octofitter_orbits}
\end{figure*}

\begin{deluxetable*}{crrrrrrr}
\tablecaption{Fitted planet parameters from the \texttt{Octofitter} orbit fit, given by the marginal posterior median and 68\% credible interval. Full corner plots may be found in the appendix.}\label{tab:orbitfit}
\tablehead{
    \multicolumn{8}{c}{\textbf{Planet Parameters}} \\
    \colhead{Planet} &
    \colhead{m ($M_{\text{Jup}}$)} &
    \colhead{a (au)} &
    \colhead{e} &
    \colhead{$i$ (deg)}&
    \colhead{$\Omega$ (deg)} &
    \colhead{$\omega$ (deg)} &
    \colhead{$\theta$ (deg)}
} 
\startdata
b & $4.65^{+2.80}_{-2.49}$ & $68.79^{+1.63}_{-3.64}$ & $0.04^{+0.05}_{-0.03}$ & $25.28^{+2.54}_{-4.42}$ & $64.29^{+7.94}_{-8.00}$ & $178.54^{+50.31}_{-48.52}$ & $55.37^{+0.10}_{-0.11}$ \\c & $9.02^{+2.91}_{-2.94}$ & $43.44^{+1.40}_{-1.31}$ & $0.02^{+0.02}_{-0.01}$ & $28.00^{+2.68}_{-3.17}$ & $63.43^{+4.30}_{-5.83}$ & $68.82^{+63.88}_{-53.95}$ & $300.58^{+0.18}_{-0.18}$ \\d & $9.17^{+2.84}_{-2.84}$ & $25.85^{+1.74}_{-1.59}$ & $0.10^{+0.07}_{-0.07}$ & $28.20^{+4.64}_{-6.24}$ & $65.36^{+5.89}_{-6.54}$ & $5.54^{+23.65}_{-19.84}$ & $165.00^{+0.82}_{-0.80}$ \\e & $11.67^{+1.87}_{-1.78}$ & $15.80^{+0.61}_{-0.50}$ & $0.13^{+0.02}_{-0.02}$ & $22.38^{+3.17}_{-3.66}$ & $58.43^{+8.18}_{-9.77}$ & $104.81^{+12.58}_{-9.88}$ & $234.99^{+0.34}_{-0.35}$ \\f & $1.65^{+1.27}_{-0.81}$ & $6.94^{+1.16}_{-0.95}$ & $0.11^{+0.18}_{-0.08}$ & $27.07^{+6.59}_{-6.52}$ & $63.69^{+13.54}_{-13.93}$ & $97.43^{+60.02}_{-54.30}$ & $354.56^{+6.99}_{-7.14}$ \\
\enddata
\end{deluxetable*}

\begin{figure*}
\centering
\includegraphics[width=0.98\textwidth]{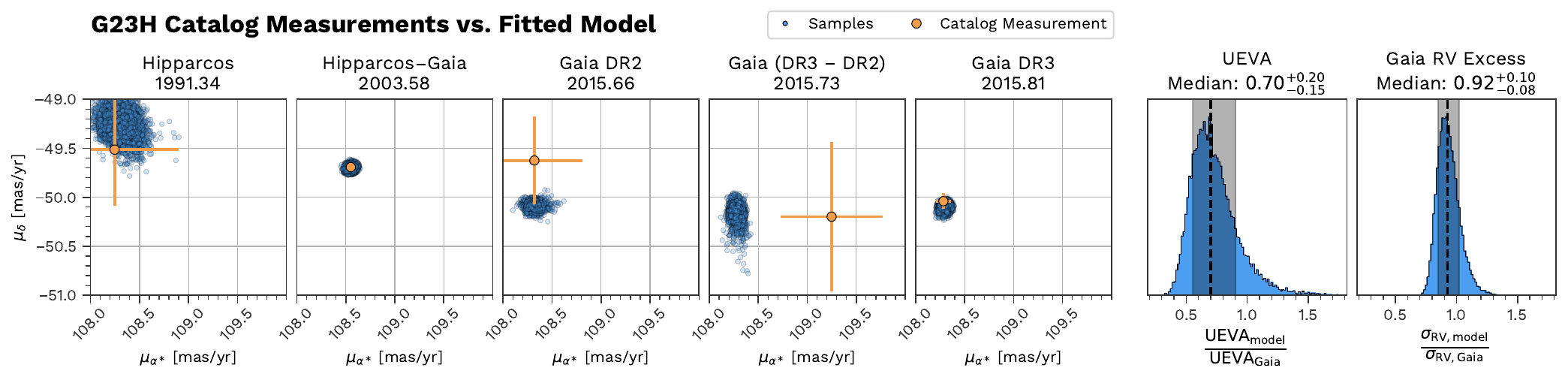}
\caption{Simulated proper motion measurements at the various Hipparcos and Gaia epochs for the fitted five-planet model compared to the actual measurements from Hipparcos and Gaia on the left five plots. The right two plots are posterior sample draws calculating the ratio of simulated measurements to measured values of UEVA and Gaia radial velocity excess, a value of 1 means that the sample matches the measurement. The median is displayed by a dashed line and the shaded region represents a 68\% credible interval on the histogram. The $1\sigma$ error bars portrayed on the five left plots do not capture the significant covariance between the $\mu_\alpha*$ and $\mu_\delta$ measurements and inter-mission epochs. Note that $\mu_\alpha*$ indicates the projected proper motion with the extra $\cos \delta$ factor applied.}\label{fig:orbit_fit_samples_vs_gh}
\end{figure*}

We present the orbit fit in Figure \ref{fig:hr_8799_octofitter_orbits} and the median value with 68\% credible interval of the marginal posterior distribution for each planetary orbit fit parameter in Table \ref{tab:orbitfit}. The full corner plots for each of the planets and candidate may be found in the appendix, Figures \ref{fig:corner_plot_b} -- \ref{fig:corner_plot_f}. Planet b has a mass most similar to the prior, while the fitted orbits for the other planets favor heavier masses than the median of the priors. The masses found are also greater than the comparable four-planet model determined by \citet{zurloOrbitalDynamicalAnalysis2022}. The median mass for candidate~f is surprisingly low, at around $1.7\,M_\text{Jup}$, with a long tail posterior and 68\% credible interval extending to $\sim3\,M_\text{Jup}$. Previous stability analyses of the system suggest a $7-9\,M_\text{Jup}$ planet while Gaia-Hipparcos astrometric analysis \citep{brandtFirstDynamicalMass2021} suggests a $3\,\sigma$ upper limit on the mass closer to $\sim 6\,M_\text{Jup}$ at the semi-major axis we measure for candidate~f. We interpret the disagreement of our fitted mass with previous analyses as evidence that the single astrometric data point for candidate~f provides insufficient data to dynamically determine the mass even when combining with absolute astrometry.

For the other orbital parameters, we find that planets b and c prefer low eccentricities ($e < 0.05$), while planets d, e, and candidate planet~f prefer eccentricities close to or higher than 0.1. The semi-major axis median posterior values for the known planets are consistent with previous estimations, and the fitted semi-major axis for candidate~f is $\sim 7$\,au. The orbits of all the planets and candidate remain roughly coplanar to within 5 degrees of $i$ and $\Omega$. The shared jitter posterior for all relative astrometry converged to a nearly Gaussian distribution, with a median value of $2.81\pm0.21$ mas. At the system level, our orbit fits yield a host star mass of $1.52\pm0.05\,M_\odot$. This fitted mass is consistent with both stellar-evolution-derived mass estimations and dynamical mass estimations. Lastly, the parallax posterior remains unchanged from the prior and we measure barycentric proper motion measurements of $\mu_{\alpha*} = 108.29\pm0.02$ mas/yr and $\mu_\delta = -50.10\pm0.03$ mas/yr, consistent with measurements from Gaia DR3.

From our orbit fits, we sample the posteriors to simulate proper motion measurements from Gaia and Hipparcos. Figure \ref{fig:orbit_fit_samples_vs_gh} shows simulated measurements (derived from gravitational force calculations from orbital posterior draws) of the measured proper motion of HR~8799 at epochs corresponding to their respective Gaia and Hipparcos missions. The figure also plots simulated measurements of the derived unbiased estimator of variance a posteriori (UEVA) \citep{kieferSearchingSubstellarCompanion2025} and Gaia radial velocity (RV) excess as compared to the Gaia DR3 values. The UEVA is consistent to within a 95\% credible interval, and the Gaia RV excess is consistent to within a 68\% credible interval.

Note that there are two inter-mission epochs, Hipparcos-Gaia and Gaia (DR3 -- DR2), which are derived proper motion measurements based on the positional differences between measurement epochs. Due to specifics on how proper motion is measured from the two missions and how the inter-mission measurements are derived, the error on $\mu_\alpha$ and $\mu_\delta$ for our simulated posterior samples and real measurements have significant covariance between each value. Accounting for this covariance, we calculate the median chi-squared of our orbit fits, yielding $\chi^2 = 10.534$. For the 10-dimensional proper motion measurements, this value is consistent with the posterior adequately describing the underlying distribution. We additionally calculate the posterior-predictive p-value, finding $p=0.384$, indicating no significant discrepancy between our model posteriors and actual measured values. While these statistical measures are decidedly self-referential (e.g. we use the Gaia and Hipparcos measurements as priors on our orbit fit), it is a good consistency check on the validity of our orbit fits and allow us to say that our model fits are consistent with measured Gaia and Hipparcos data.


\subsection{Derived Planet Photometry}

\begin{figure}
\centering
\includegraphics[width=0.49\textwidth]{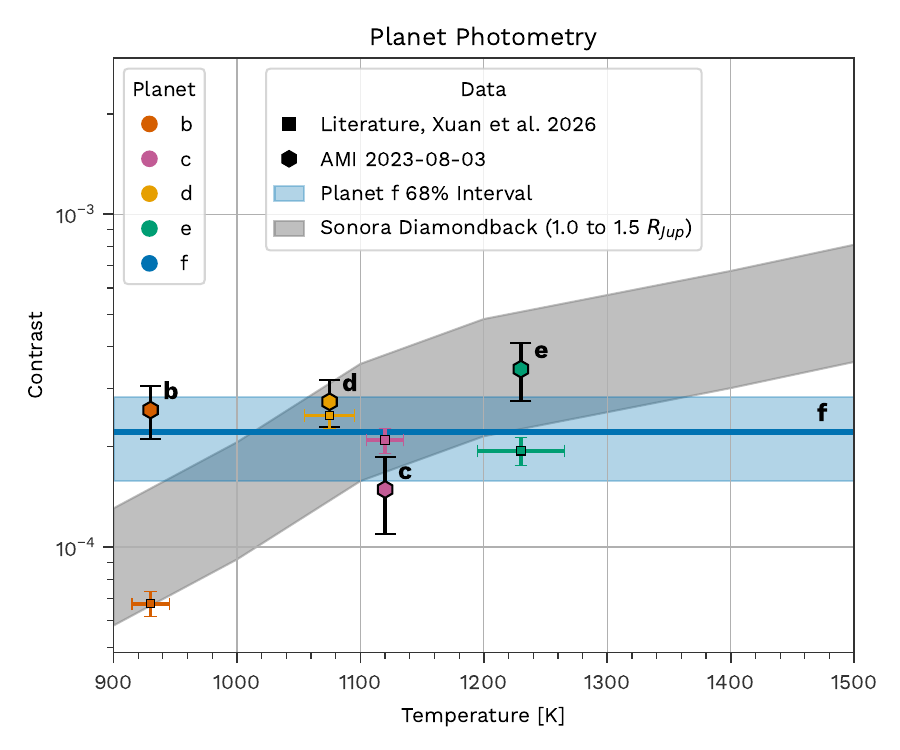}
\caption{Recovered JWST \ac{ami} contrast for all 4 planets and the candidate planet in the system as compared to the bolometric luminosity from \cite{xuanCompositionsHR87992026} integrated over the NIRISS F380M bandpass. We use the retrieved temperatures for the 4 known planets and do not assume any temperature for candidate planet f. The blue line indicates the median photometry for candidate planet f, and the shaded blue region is its 68\% credible interval on the posterior. The gray shaded region is the derived contrast from the stellar and planetary atmosphere models for a $1.0$--$1.5\,R_\text{Jup}$ planet.}\label{fig:photometry_plot}
\end{figure}

\begin{deluxetable}{cccc}
\tablecaption{Derived F380M contrasts from estimates of the planet bolometric luminosity and the contrasts directly measured with AMIGO.}
\label{tab:contrasts}
\tablehead{
    \colhead{Planet} &
    \colhead{$T_{\mathrm{eff}}$ (K)} &
    \colhead{$C_{\mathrm{Xuan}}$ ($\times10^{-4}$)} &
    \colhead{$C_{\mathrm{AMIGO}}$ ($\times10^{-4}$)}
}
\startdata
b & 930 & $0.68\pm0.06$ & $2.58\pm0.47$ \\
c & 1120 & $2.10\pm0.18$ & $1.49\pm0.39$ \\
d & 1075 & $2.48\pm0.21$ & $2.73\pm0.44$ \\
e & 1230 & $1.95\pm0.19$ & $3.43\pm0.68$ \\
f & -- & -- & $2.21\pm0.62$
\enddata
\end{deluxetable}

We compare our \ac{ami} photometry to literature values from \cite{xuanCompositionsHR87992026} by remapping the bolometric luminosities found from their retrieved models to \ac{niriss} \ac{ami} F380M measurements. This rescaling includes using the overall photon-to-electron conversion efficiency curves for the F380M filter. To obtain these measurements, we use the \texttt{PHOENIX} stellar atmosphere models \citep{husserNewExtensiveLibrary2013a} and the \texttt{Sonora Diamondback} \citep{morleySonoraSubstellarAtmosphere2024} substellar atmosphere models, noting that previous atmospheric analyses favor cloudy models \citep{madhusudhanMODELATMOSPHERESMASSIVE2011, nasedkinFourofakindComprehensiveAtmospheric2024a}. To perform a simple photometric comparison, we assume the following parameters for all 4 planets: metallicity $[M/H]=+0.5$, $\text{C/O}=0.458$ (solar), surface gravity $\log g=4$, and cloud parameterization $f_\text{sed}=2$, where these are the nearest values to the latest retrievals performed. For simplicity, we did not vary $f_\text{sed}$, but changing this parameter did not significantly change the results. We use temperature values closest to the retrieved planet temperature when integrating over the F380M bandpass. Lastly, we use a stellar radius of $R_\star = 1.44\pm0.06\, R_\odot$ \citep{bainesCHARAARRAYANGULAR2012}. The errors propagated from this term serve as the largest uncertainties on the derived contrasts. The resulting contrasts we compute are shown in Figure \ref{fig:photometry_plot} and are listed in Table \ref{tab:contrasts}. Our recovered photometry appears to be most consistent within a 68\% credible interval for planet d, consistent to within a 95\% credible interval for planets c and e, and are inconsistent with planet b. This inconsistency will be discussed in the following section.

The limited wavelength coverage limits photometric interpretation of candidate~f. With only a single photometric bandpass, the properties of the candidate are difficult to deduce. Future observations may cover \ac{niriss} \ac{ami} filters F380M, F430M, and F480M with subpixel dithering, allowing us to begin to constrain the spectral energy distribution of candidate~f.

\subsection{Astrometric \& Photometric Consistency} \label{sec:consistency}

The recovered properties of planets bcde are broadly consistent with expectations but some individual astrometric and photometric measurements show discrepancies from well known properties. These differences may reflect a combination of several sources of systematics that are difficult to quantify in this study. In particular, the GTO~1200 data only consists of one dither that makes it more difficult to separate the detector response from the astrophysical signal. The detector model in \ac{amigo} is trained using a separate calibration dataset with its parameters held fixed throughout our analysis. The inferred companion properties may therefore retain some sensitivity to uncertainties in this calibration. Residual uncertainty in flat-fielding calibration are also a possible contribution, since imperfect correction of inter-pixel sensitivities could subtly alter the model intensities across the image. The single dither combined with detector calibration errors may manifest as residuals in the image-plane that can bias inferred positions and contrasts depending on the companion location. Additional calibration data and observations at multiple subpixel dither positions would help constrain these contributions and assess their impact on the recovered companion properties. The demonstrated performance of \ac{amigo} on benchmark systems however, supports its effectiveness in recovering planet properties. The issues discussed here are small, but relevant at the precision of our measurements, thus leaving room for refinement in calibration. Follow up observations in multiple filters that better sample subpixel positions would mitigate some of these issues.

\section{Discussion \& Conclusion}\label{sec:disc}

HR~8799 is one of the most well-studied exoplanet systems and is a defining example of the science of direct imaging of exoplanets. As we continue to study this system, we find that it is increasingly unique amongst our population of known exoplanet systems. If confirmed, candidate planet~f would mark the fifth planet discovered in HR~8799, revealing an even richer planetary architecture that challenges our current understanding of planet formation and evolution in the system.

By combining relative astrometry and absolute astrometry from Gaia and Hipparcos, we better constrain the orbit of HR~8799 than a fit with relative astrometry alone. While we were unable to dynamically determine the mass of candidate~f given the data from the \ac{ami} epoch, the imminent release of Gaia DR4 will put significantly better constraints on the stellar reflex motion of the system, thus allowing us to constrain the dynamics of the system to even higher accuracy. Continued monitoring is essential for refining our understanding of the system, as its dynamical evolution provides key constraints on fundamental properties, including the masses of the host star and planets. A comprehensive dynamic analysis of the system will need to jointly model of all aspects of the system, encompassing absolute astrometry, relative astrometry, and radial velocity measurements. Parameters such as the parallax, proper motion, and host star mass should be fitted alongside the orbital parameters with the appropriate priors to ensure uncertainties and covariances are properly propagated into orbital parameters. We hope that this analysis provides a framework for future dynamical studies of the HR~8799 system as additional astrometry and radial velocity measurements become available.

Future dynamical studies could also test whether candidate~f is truly in resonance at its measured position. When comparing the resonant Va model predicted location of planet~f from \cite{gozdziewskiMultipleMeanMotion2014a} at the epoch of the \ac{ami} observation, we find that the predicted location of the planet is quite close to our recovered astrometry, despite their model only using astrometric data up until 2011. Notably, the more recent WITP orbital predictions trail the Va model predictions as seen in Figure \ref{fig:hr_8799_octofitter_orbits}, likely due to the more recent astrometry. While our data point exists near a known five-planet resonant solution, we do not claim this as direct evidence that our candidate is in resonance. The remaining resonant models Vb, Vc, and Vd, all have planet~f at various locations around the star azimuthally but at projected separations similar to that of our candidate, so agreement could simply be coincidental. However, a resonant configuration may help maintain long-term dynamical stability in the system. Nevertheless, the candidate's measured position is near at least one resonant solution, and issues of resonance may be elaborated on in future studies.


Previous searches provide several constraints on the interpretation of candidate~f. \cite{wahhajSearchFifthPlanet2021} places contrasts limits at shorter near infrared wavelengths (YJHK-bands). While their contrast limits are compelling, a possible explanation for their non-detection of our candidate may be that candidate~f lies at a contrast below their calculated limits at these shorter wavelengths. At longer wavelengths, where the planet contrast can be more favorable for detection, the deep L'-band search by \cite{thompsonDeepOrbitalSearch2022} reports a candidate innermost planet with a comparable contrast to our \ac{ami} candidate~f. This is a particularly relevant comparison as their observations cover a similar wavelength range to \ac{niriss} F380M, recovering a source with a semi-major axis of $4-5$\,au. However, our candidate's astrometry appears inconsistent with the position predicted by their orbital posterior at our observing epoch, suggesting that the two signals are unlikely to represent the same object. We cannot presently explain the absence of a corresponding source in \cite{thompsonDeepOrbitalSearch2022}, and acknowledge that this non-detection remains in potential tension with our candidate. We emphasize that evidence for our candidate planet~f remains tentative and future observations are necessary to confirm it.

Presently, this measurement is only known to be possible with \ac{niriss} \ac{ami} on JWST. Observations of HR~8799 using NIRSpec \citep{ruffioJupiterlikeUniformMetal2026, xuanCompositionsHR87992026} and NIRCam \citep{Balmer_2025} are not currently sensitive to candidate planet~f at the separation and contrast we measure. However, this object may visible from the ground using the newly upgraded GRAVITY+ \citep{abuterFirstLightGRAVITY2026} as this object is within their sensitivity.

The detection of candidate planet~f would not have been possible without the combination of the capabilities of \ac{niriss} \ac{ami} combined with the novel end-to-end forward modeling code \ac{amigo}. Follow up observations with \ac{niriss} \ac{ami} in multiple filters, increased group count, and subpixel dithers will be critical to confirming this object. Significant progress in optimal \ac{ami} observing settings to combat \ac{bfe} and other systematics \citep{Sallum2023, desdoigtsAmigoDatadrivenCalibration2026} will also allow us to obtain an even more ideal dataset for this system. Furthermore, additional filters will improve the robustness of this measurement and the time since the initial observation will allow us to measure common proper motion and definitively confirm the existence of the object. 

The differentiable forward-modeling paradigm has extended our sensitivity to a new region of parameter space at higher contrasts and lower separation that was previously unavailable to current technology. Development is underway for an update to \ac{amigo} that refactors the detector model, including updated calibration training data that may improve performance and reduce systematics. This and future updates will push the sensitivity of \ac{niriss} \ac{ami} to higher accuracy and precision, potentially refining the inferred properties the candidate planet~f through analyses of archival and future data. We are now realizing the potential of \ac{ami} with forward modeling tools, reaching high contrast at small angular separations, allowing us to probe for faint companions in the innermost regions around nearby stars at solar system scales.

\begin{acknowledgments}
This work is based in part on observations made with the NASA/ESA/CSA James Webb Space Telescope. The data were obtained from the Mikulski Archive for Space Telescopes at the Space Telescope Science Institute, which is operated by the Association of Universities for Research in Astronomy, Inc., under NASA contract NAS5-03127 for JWST. These observations are associated with program GTO~1200. The authors acknowledge the PI T.V. and co-PIs for developing their observing program with a zero-exclusive-access period. 

This research has made use of the Keck Observatory Archive (KOA), which is operated by the W. M. Keck Observatory and the NASA Exoplanet Science Institute (NExScI), under contract with the National Aeronautics and Space Administration. The authors wish to recognize and acknowledge the very significant cultural role and reverence that the summit of Maunakea has always had within the Native Hawaiian community. We are most fortunate to have the opportunity to conduct observations from this mountain.

BP and PT have been supported by the Australian Research Council grant DP230101439; and MC and LD have been supported by the Australian Government Research Training Program (RTP) award. We are grateful to the Australian public for enabling this science. Development of \dlux has been supported by the Breakthrough Foundation through their Toliman project as a part of the Breakthrough Watch initiative. 

BP, MC, and PT acknowledge and pay respect to the traditional owners of the land on which the University of Sydney and Macquarie University are situated, upon whose unceded, sovereign, ancestral lands we work. We pay respects to their Ancestors and descendants, who continue cultural and spiritual connections to Country. 

DJ is supported by NRC Canada and by an NSERC Discovery Grant.

This work was co-authored by an employee of Caltech/IPAC under Contract No. 80GSFC21R0032 with the National Aeronautics and Space Administration.

\end{acknowledgments}

\begin{contribution}
J.S.N. led the data reduction, data analysis, and manuscript preparation. L.D.,  M.C., and B.J.S.P. developed the data modeling package \ac{amigo}, including the scripts that were used to reduce and analyze data. B.J.S.P. wrote the methods portion of the manuscript describing \ac{amigo}. A.Z.G. provided significant direction for the project and manuscript. K.H. wrote the scripts used to fit the orbits using \texttt{Octofitter}. All authors contributed scientific expertise in the interpretation of data and analysis presented.
\end{contribution}

\facilities{JWST(NIRISS), Keck II(NIRC2)}

\software{\ac{amigo} \citep{desdoigtsAmigoDatadrivenCalibration2026}, \dlux \citep{dlux1, dlux2}, \texttt{Octofitter} \citep{Octofitter}, \texttt{ADI.jl} \citep{lucasADIjlJuliaPackage2020}, \texttt{whereistheplanet} \citep{wangWhereistheplanetPredictingPositions2021},  \citep{astropy:2013, astropy:2018, astropy:2022}, \texttt{SciPy} \citep{scipy}, \texttt{NumPy} \citep{numpy}, \texttt{Matplotlib} \citep{matplotlib}, \texttt{numpyro} \citep{numpyro_1,numpyro_2}, \texttt{jax} \citep{jax}}

\appendix

\section{2025 October 09 NIRC2 Keck Dataset}

\begin{deluxetable}{crr}[h]
\tablecaption{Recovered astrometry and photometry from the 2025 October 09 NIRC2 Keck dataset.} \label{tab:kecknirc2}
\tablehead{
    \colhead{Planet} &
    \colhead{$\Delta$RA (mas)} &
    \colhead{$\Delta$Dec (mas)}
}
\startdata
b & $1641.8 \pm 4.6$ & $477.8 \pm 4.3$ \\
c & $-223.5 \pm 2.9$ & $928.0 \pm 2.8$ \\
d & $-635.6 \pm 3.5$ & $-278.1 \pm 2.7$ \\
e & $-147.0 \pm 8.1$ & $372.4 \pm 10.2$ \\
\enddata
\end{deluxetable}

We observed HR~8799 on 2025 October 09, using NIRC2 on Keck II the W.M. Keck Observatory. Total integration time was 1716 seconds in L'-band and no dithers were performed. The target was observed through transit totalling 166~degrees of rotation. Seeing was 0.45" as measured by the DIMM at Mauna Kea weather center. The data were reduced with standard data reduction techniques, subtracting darks and flat fielding using sky flats. For post-processing, we used the \texttt{ADI.jl} package \citep{lucasADIjlJuliaPackage2020}, subtracting the host star PSF using the first 20~PCA modes. All 4 known planets were clearly recovered in the reduction. The planet locations were then obtained using the cross correlation of a forward modeled template PSF.

\section{Corner Plots}

\begin{figure}
\centering
\includegraphics[width=0.97\textwidth]{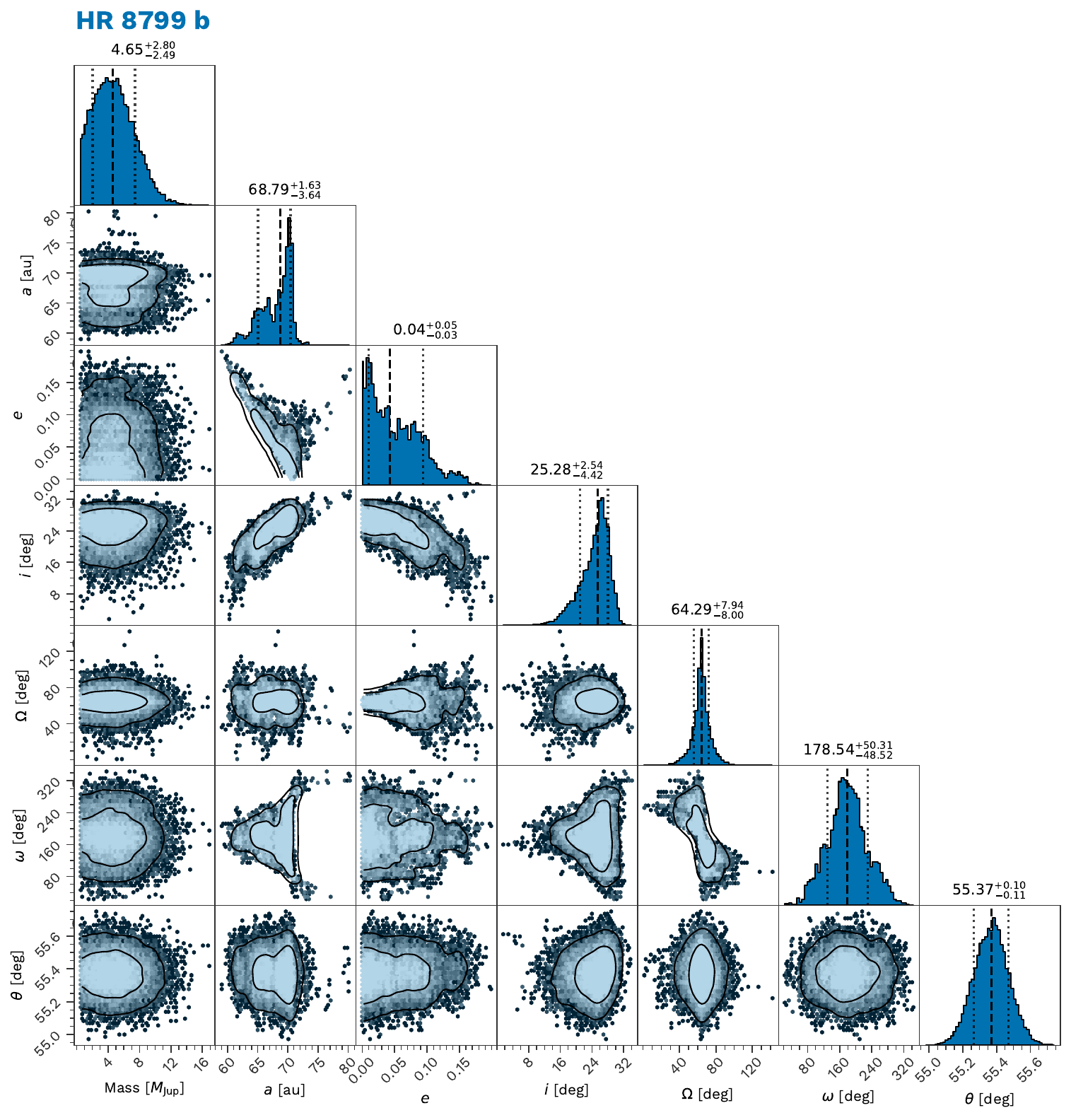}
\caption{Posterior distributions from the orbit fit for a subset of the full model parameters for planet b. Distributions are shown with 68\% and 95\% credible region contour outlines and the normalized histogram shows the median and $\pm1\,\sigma$ values. Sample draws are binned onto a hexagonal grid with the color plotted on a log scale. Each hexagonal histogram bin has a minimum value of 1 count per bin.}\label{fig:corner_plot_b}
\end{figure}

\begin{figure}
\centering
\includegraphics[width=0.97\textwidth]{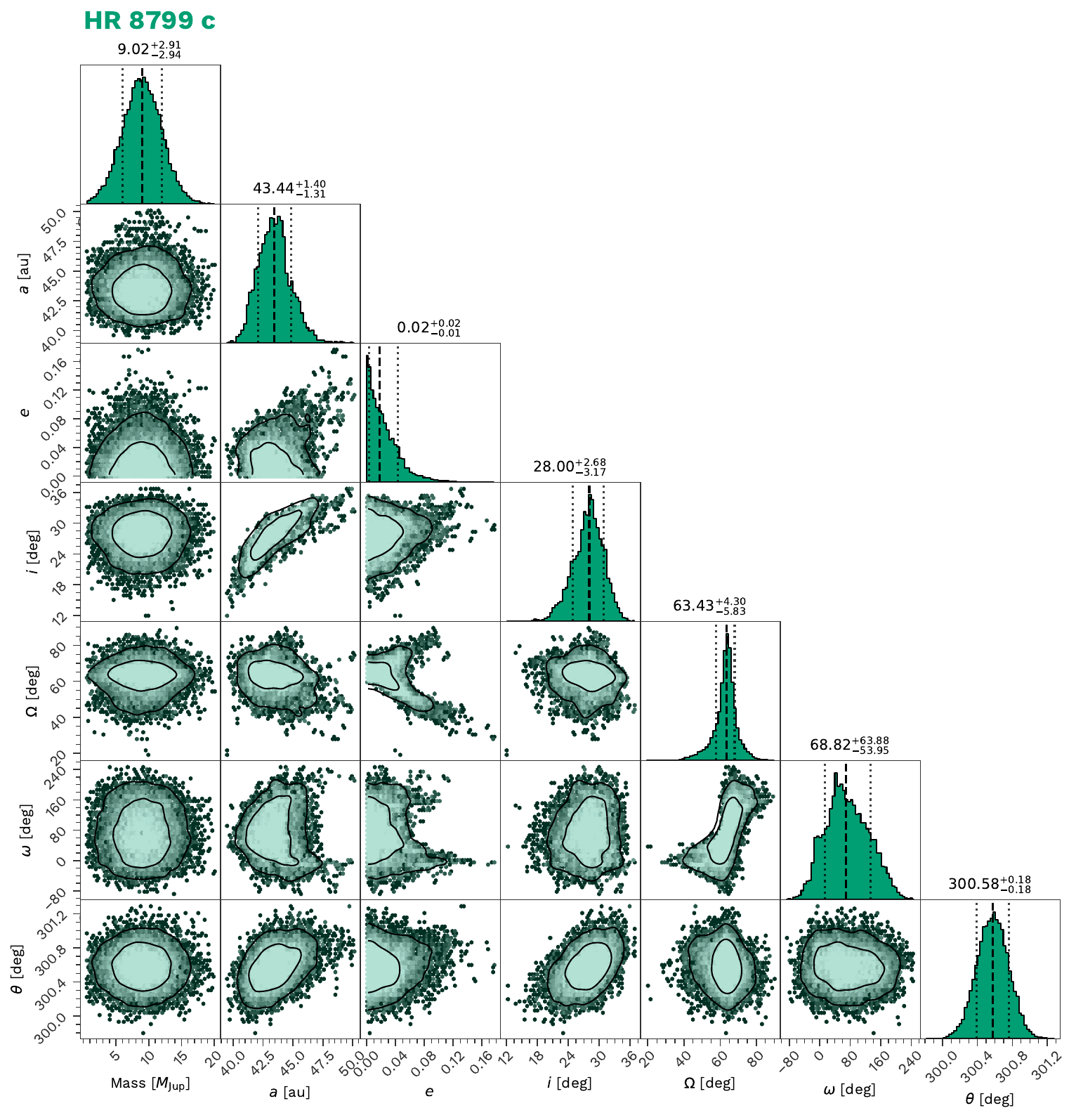}
\caption{Posterior distributions from the orbit fit for a subset of the full model parameters for planet c. Distributions are shown with 68\% and 95\% credible region contour outlines and the normalized histogram shows the median and $\pm1\,\sigma$ values. Sample draws are binned onto a hexagonal grid with the color plotted on a log scale. Each hexagonal histogram bin has a minimum value of 1 count per bin.}\label{fig:corner_plot_c}
\end{figure}

\begin{figure}
\centering
\includegraphics[width=0.97\textwidth]{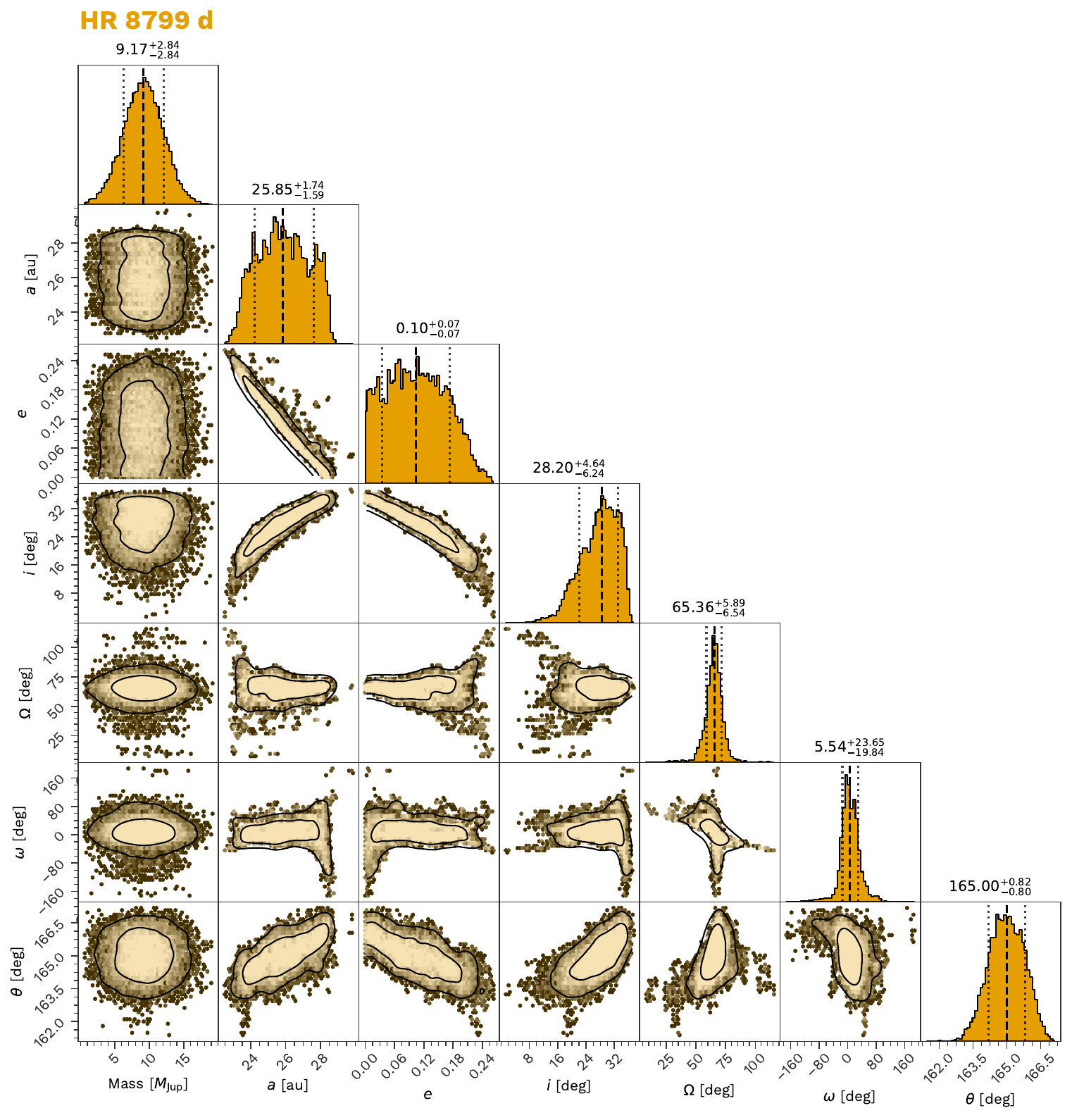}
\caption{Posterior distributions from the orbit fit for a subset of the full model parameters for planet d. Distributions are shown with 68\% and 95\% credible region contour outlines and the normalized histogram shows the median and $\pm1\,\sigma$ values. Sample draws are binned onto a hexagonal grid with the color plotted on a log scale. Each hexagonal histogram bin has a minimum value of 1 count per bin.}\label{fig:corner_plot_d}
\end{figure}

\begin{figure}
\centering
\includegraphics[width=0.97\textwidth]{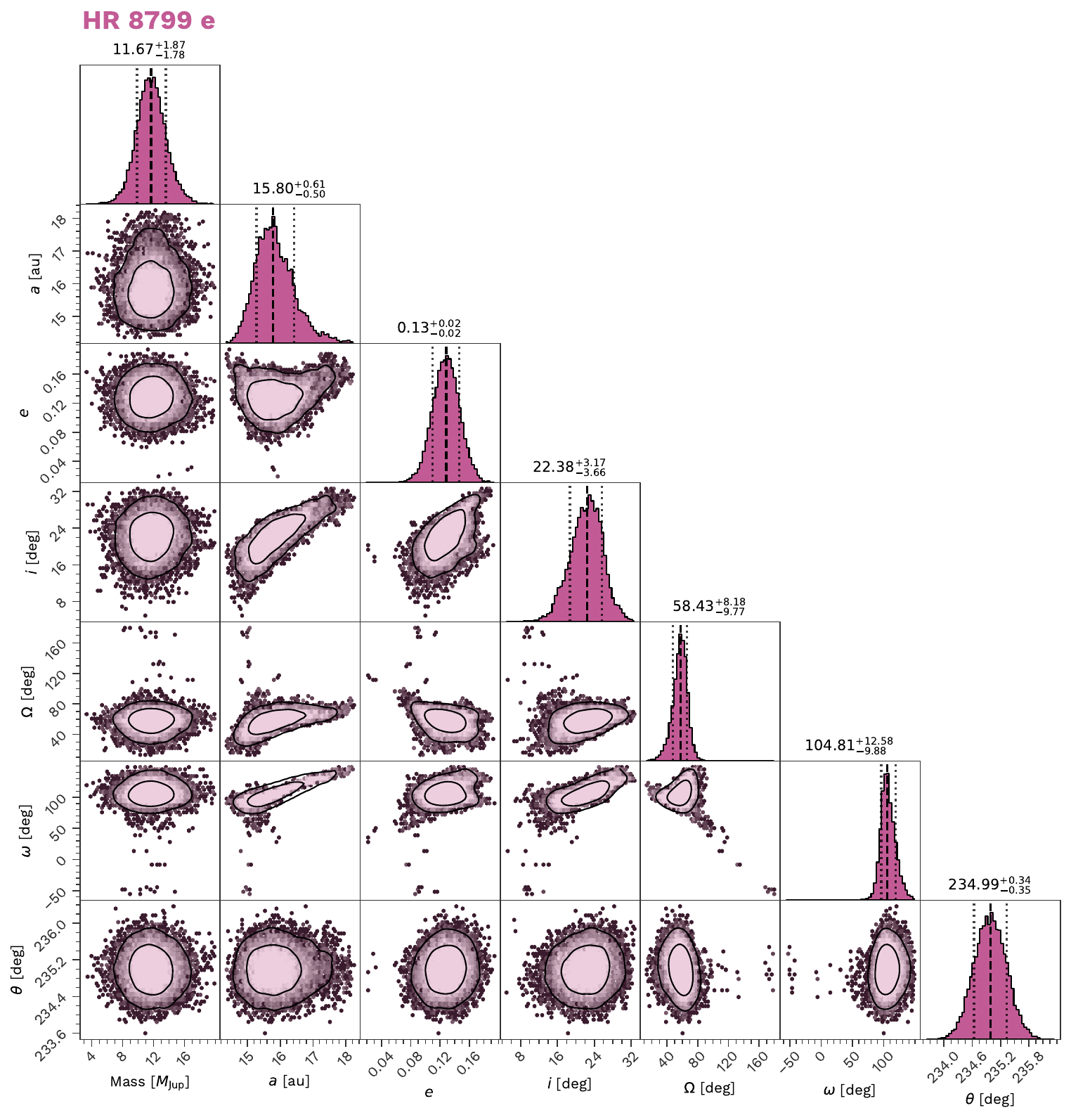}
\caption{Posterior distributions from the orbit fit for a subset of the full model parameters for planet e. Distributions are shown with 68\% and 95\% credible region contour outlines and the normalized histogram shows the median and $\pm1\,\sigma$ values. Sample draws are binned onto a hexagonal grid with the color plotted on a log scale. Each hexagonal histogram bin has a minimum value of 1 count per bin.}\label{fig:corner_plot_e}
\end{figure}

\begin{figure}
\centering
\includegraphics[width=0.97\textwidth]{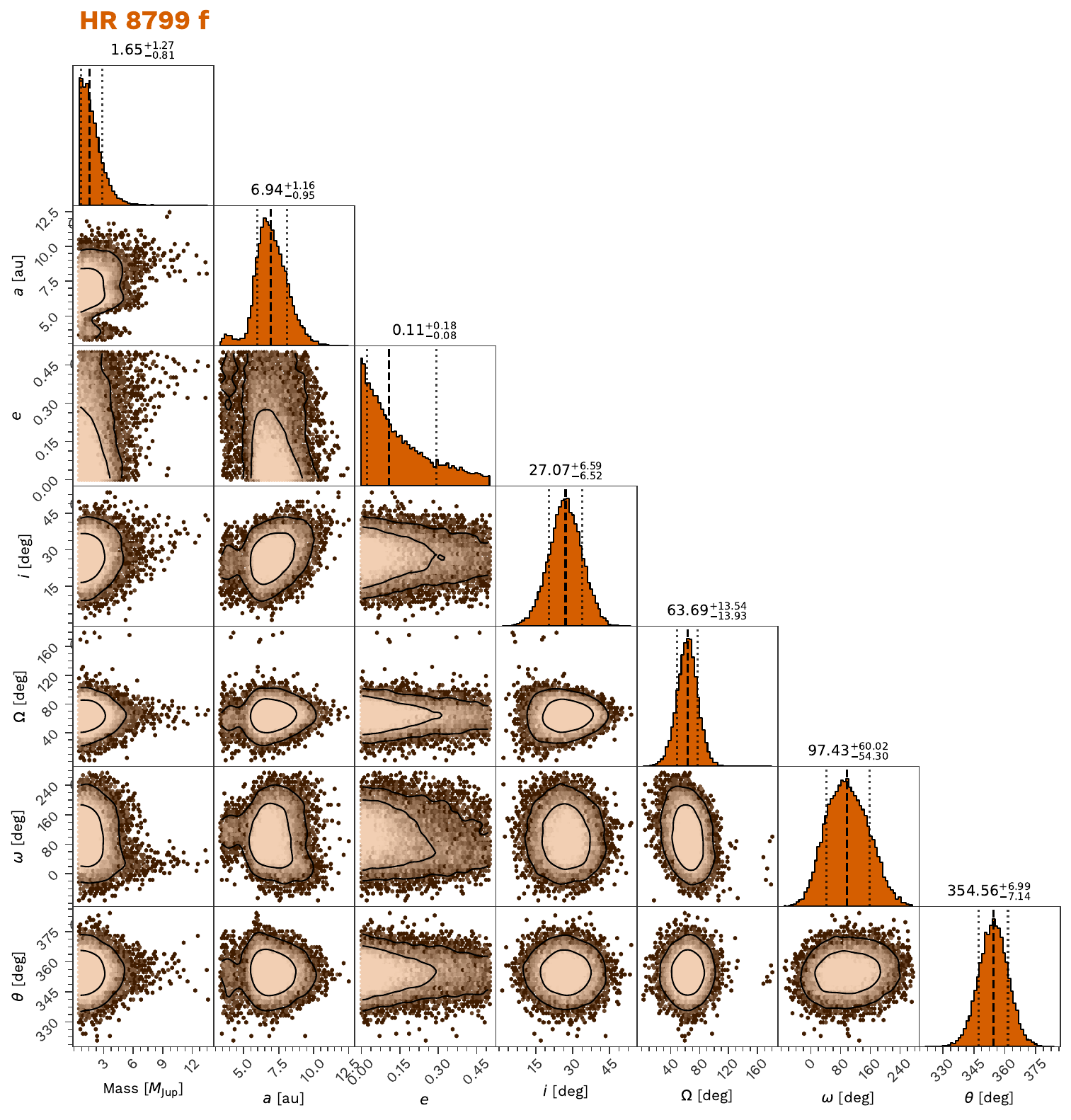}
\caption{Posterior distributions from the orbit fit for a subset of the full model parameters for candidate planet~f. Distributions are shown with 68\% and 95\% credible region contour outlines and the normalized histogram shows the median and $\pm1\sigma$ values. Sample draws are binned onto a hexagonal grid with the color plotted on a log scale. Each hexagonal histogram bin has a minimum value of 1 count per bin.}\label{fig:corner_plot_f}
\end{figure}

\clearpage
\bibliography{bibliography}{}

@ARTICLE{closure_phase,
       author = {{Jennison}, R.~C.},
        title = "{A phase sensitive interferometer technique for the measurement of the Fourier transforms of spatial brightness distributions of small angular extent}",
      journal = {\mnras},
         year = 1958,
        month = jan,
       volume = {118},
        pages = {276},
          doi = {10.1093/mnras/118.3.276},
       adsurl = {https://ui.adsabs.harvard.edu/abs/1958MNRAS.118..276J}
}

@ARTICLE{Martinache2010,
       author = {{Martinache}, Frantz},
        title = "{Kernel Phase in Fizeau Interferometry}",
      journal = {\apj},
         year = 2010,
        month = nov,
       volume = {724},
       number = {1},
        pages = {464-469},
          doi = {10.1088/0004-637X/724/1/464},
archivePrefix = {arXiv},
       eprint = {1009.3933},
 primaryClass = {astro-ph.IM},
       adsurl = {https://ui.adsabs.harvard.edu/abs/2010ApJ...724..464M}
}

@ARTICLE{Pope2016,
       author = {{Pope}, Benjamin J.~S.},
        title = "{Kernel phase and kernel amplitude in Fizeau imaging}",
      journal = {\mnras},
         year = 2016,
        month = dec,
       volume = {463},
       number = {4},
        pages = {3573-3581},
          doi = {10.1093/mnras/stw2215},
archivePrefix = {arXiv},
       eprint = {1609.00200},
 primaryClass = {astro-ph.IM},
       adsurl = {https://ui.adsabs.harvard.edu/abs/2016MNRAS.463.3573P}
}

@inproceedings{niriss1,
	title        = {The JWST Fine Guidance Sensor (FGS) and Near-Infrared Imager and Slitless Spectrograph (NIRISS)},
	author       = {Ren{\'e} Doyon and John B. Hutchings and Mathilde Beaulieu and Loic Albert and David Lafreni{\`e}re and Chris Willott and Driss Touahri and Neil Rowlands and Micheal Maszkiewicz and Alex W. Fullerton and Kevin Volk and Andr{\'e} R. Martel and Pierre Chayer and Anand Sivaramakrishnan and Roberto Abraham and Laura Ferrarese and Ray Jayawardhana and Doug Johnstone and Michael Meyer and Judith L. Pipher and Marcin Sawicki},
	year         = 2012,
	booktitle    = {Space Telescopes and Instrumentation 2012: Optical, Infrared, and Millimeter Wave},
	publisher    = {SPIE},
	volume       = 8442,
	pages        = {84422R},
	doi          = {10.1117/12.926578},
	url          = {https://doi.org/10.1117/12.926578},
	editor       = {Mark C. Clampin and Giovanni G. Fazio and Howard A. MacEwen and Jacobus M. Oschmann Jr.},
	organization = {International Society for Optics and Photonics},
}

@article{niriss2,
  title = {The {{Near Infrared Imager}} and {{Slitless Spectrograph}} for the {{James Webb Space Telescope}}. {{I}}. {{Instrument Overview}} and {{In-flight Performance}}},
  author = {Doyon, Ren{\'e} and Willott, Chris J. and Hutchings, John B. and Sivaramakrishnan, Anand and Albert, Lo{\"i}c and Lafreni{\`e}re, David and Rowlands, Neil and Bego{\~n}a Vila, M. and Martel, Andr{\'e} R. and LaMassa, Stephanie and Aldridge, David and Artigau, {\'E}tienne and Cameron, Peter and Chayer, Pierre and Cook, Neil J. and Cooper, Rachel A. and {Darveau-Bernier}, Antoine and Dupuis, Jean and Earnshaw, Colin and Espinoza, N{\'e}stor and Filippazzo, Joseph C. and Fullerton, Alexander W. and Gaudreau, Daniel and Gawlik, Roman and Goudfrooij, Paul and Haley, Craig and Kammerer, Jens and Kendall, David and Lambros, Scott D. and Ignat, Luminita Ilinca and Maszkiewicz, Michael and McColgan, Ashley and Morishita, Takahiro and Ouellette, Nathalie N.-Q. and Pacifici, Camilla and Philippi, Natasha and Radica, Michael and Ravindranath, Swara and Rowe, Jason and Roy, Arpita and Roy, Niladri and Saad, Karl and Sohn, Sangmo Tony and Talens, Geert Jan and Touahri, Driss and Thatte, Deepashri and Taylor, Joanna M. and Vandal, Thomas and Volk, Kevin and Wander, Michel and Warner, Gerald and Zheng, Sheng-Hai and Zhou, Julia and Abraham, Roberto and Beaulieu, Mathilde and Benneke, Bj{\"o}rn and Ferrarese, Laura and Jayawardhana, Ray and Johnstone, Doug and Kaltenegger, Lisa and Meyer, Michael R. and Pipher, Judy L. and Rameau, Julien and Rieke, Marcia and Salhi, Salma and Sawicki, Marcin},
  year = 2023,
  month = sep,
  journal = {Publications of the Astronomical Society of the Pacific},
  volume = {135},
  number = {1051},
  pages = {098001},
  publisher = {The Astronomical Society of the Pacific},
  issn = {1538-3873},
  doi = {10.1088/1538-3873/acd41b},
  urldate = {2026-09-03},
  langid = {english}
}

@article{bfe_miri,
	title        = {The brighter-fatter effect in the JWST MIRI Si:As IBC detectors: I. Observations, impact on science, and modeling},
	author       = {Argyriou, Ioannis and Lage, Craig and Rieke, George H. and Gasman, Danny and Bouwman, Jeroen and Morrison, Jane and Libralato, Mattia and Dicken, Daniel and Brandl, Bernhard R. and \'{A}lvarez-M\'{a}rquez, Javier and Labiano, Alvaro and Regan, Michael and Ressler, Michael E.},
	year         = 2023,
	month        = dec,
	journal      = {\aap},
	publisher    = {EDP Sciences},
	volume       = 680,
	pages        = {A96},
	doi          = {10.1051/0004-6361/202346490},
	issn         = {1432-0746},
	url          = {http://dx.doi.org/10.1051/0004-6361/202346490},
}

@ARTICLE{Choi2020,
       author = {{Choi}, Ami and {Hirata}, Christopher M.},
        title = "{Brighter-fatter Effect in Near-infrared Detectors. II. Autocorrelation Analysis of H4RG-10 Flats}",
      journal = {\pasp},
         year = 2020,
        month = jan,
       volume = {132},
       number = {1007},
          eid = {014502},
        pages = {014502},
          doi = {10.1088/1538-3873/ab4504},
archivePrefix = {arXiv},
       eprint = {1906.01847},
 primaryClass = {astro-ph.IM},
       adsurl = {https://ui.adsabs.harvard.edu/abs/2020PASP..132a4502C}
}

@ARTICLE{Hirata2020,
       author = {{Hirata}, Christopher M. and {Choi}, Ami},
        title = "{Brighter-fatter Effect in Near-infrared Detectors. I. Theory of Flat Autocorrelations}",
      journal = {\pasp},
         year = 2020,
        month = jan,
       volume = {132},
       number = {1007},
          eid = {014501},
        pages = {014501},
          doi = {10.1088/1538-3873/ab44f7},
archivePrefix = {arXiv},
       eprint = {1906.01846},
 primaryClass = {astro-ph.IM},
       adsurl = {https://ui.adsabs.harvard.edu/abs/2020PASP..132a4501H}
}

@article{blakelyJamesWebbInterferometer2025b,
  title = {The {{James Webb Interferometer}}: {{Space-based Interferometric Detections}} of {{PDS}} 70 b and c at 4.8 {$\mu$}m},
  shorttitle = {The {{James Webb Interferometer}}},
  author = {Blakely, Dori and Johnstone, Doug and Cugno, Gabriele and Sivaramakrishnan, Anand and Tuthill, Peter and Dong, Ruobing and Pope, Benjamin J. S. and Albert, Lo{\"i}c and Charles, Max and Cooper, Rachel A. and Furio, Matthew De and Desdoigts, Louis and Doyon, Ren{\'e} and Francis, Logan and Greenbaum, Alexandra Z. and Lafreni{\`e}re, David and Lloyd, James P. and Meyer, Michael R. and Pueyo, Laurent and Ray, Shrishmoy and {S{\'a}nchez-Berm{\'u}dez}, Joel and Soulain, Anthony and Thatte, Deepashri and Thompson, William and Vandal, Thomas},
  year = 2025,
  month = feb,
  journal = {The Astronomical Journal},
  volume = {169},
  number = {3},
  pages = {137},
  publisher = {The American Astronomical Society},
  issn = {1538-3881},
  doi = {10.3847/1538-3881/ad9b94},
  urldate = {2026-09-03},
  langid = {english}
}

@article{autodiff,
	title        = {A review of automatic differentiation and its efficient implementation},
	author       = {Margossian, Charles C.},
	year         = 2019,
	month        = mar,
	journal      = {WIREs Data Mining and Knowledge Discovery},
	publisher    = {Wiley},
	volume       = 9,
	number       = 4,
	doi          = {10.1002/widm.1305},
	issn         = {1942-4795},
	url          = {http://dx.doi.org/10.1002/WIDM.1305},
}

@article{dlux1,
  title = {Differentiable Optics with {$\partial$}{{Lux}}: {{I}}---Deep Calibration of Flat Field and Phase Retrieval with Automatic Differentiation},
  shorttitle = {Differentiable Optics with {$\partial$}{{Lux}}},
  author = {Desdoigts, Louis and Pope, Benjamin J. S. and Dennis, Jordan and Tuthill, Peter G.},
  year = 2023,
  month = jun,
  journal = {Journal of Astronomical Telescopes, Instruments, and Systems},
  volume = {9},
  number = {2},
  pages = {028007},
  publisher = {SPIE},
  issn = {2329-4124},
  doi = {10.1117/1.JATIS.9.2.028007},
  urldate = {2026-09-09},
  langid = {english}
}

@misc{dlux2,
	title        = {Differentiable Optics with dLux II: Optical Design Maximising Fisher Information},
	author       = {Louis Desdoigts and Benjamin Pope and Michael Gully-Santiago and Peter Tuthill},
	year         = 2024,
	url          = {https://arxiv.org/abs/2406.08704},
	eprint       = {2406.08704},
	archiveprefix = {arXiv},
	primaryclass = {astro-ph.IM},
}

@misc{scipy,
	title        = {{SciPy}: Open source scientific tools for Python},
	author       = {Jones, E. and Oliphant, T. and Peterson, P. and Others},
	year         = 2001,
	url          = {http://www.scipy.org/},
}

@article{matplotlib,
	title        = {Matplotlib: A 2D graphics environment},
	author       = {Hunter, J. D.},
	year         = 2007,
	journal      = {Computing In Science \& Engineering},
	publisher    = {IEEE COMPUTER SOC},
	volume       = 9,
	number       = 3,
	pages        = {90--95},
}

@article{numpy,
	title        = {Array programming with {NumPy}},
	author       = {Harris, Charles R. and Millman, K. Jarrod and van der Walt, St{\'e}fan J. and Gommers, Ralf and Virtanen, Pauli and Cournapeau, David and Wieser, Eric and Taylor, Julian and Berg, Sebastian and Smith, Nathaniel J. and Kern, Robert and Picus, Matti and Hoyer, Stephan and van Kerkwijk, Marten H. and Brett, Matthew and Haldane, Allan and del R{\'\i}o, Jaime Fern{\'a}ndez and Wiebe, Mark and Peterson, Pearu and G{\'e}rard-Marchant, Pierre and Sheppard, Kevin and Reddy, Tyler and Weckesser, Warren and Abbasi, Hameer and Gohlke, Christoph and Oliphant, Travis E.},
	year         = 2020,
	journal      = {Nature},
	volume       = 585,
	number       = 7825,
	pages        = {357--362},
}

@software{jax,
	title        = {{JAX}: composable transformations of {P}ython+{N}um{P}y programs},
	author       = {James Bradbury and Roy Frostig and Peter Hawkins and Matthew James Johnson and Chris Leary and Dougal Maclaurin and George Necula and Adam Paszke and Jake Vander{P}las and Skye Wanderman-{M}ilne and Qiao Zhang},
	year         = 2018,
	url          = {http://github.com/jax-ml/jax},
	version      = {0.3.13},
}

@article{ruffio_2018,
	title        = {A Bayesian Framework for Exoplanet Direct Detection and Non-detection},
	author       = {Ruffio, Jean-Baptiste and Mawet, Dimitri and Czekala, Ian and Macintosh, Bruce and De Rosa, Robert J. and Ruane, Garreth and Bottom, Michael and Pueyo, Laurent and Wang, Jason J. and Hirsch, Lea and Zhu, Zhaohuan and Nielsen, Eric L.},
	year         = 2018,
	month        = oct,
	journal      = {\aj},
	publisher    = {The American Astronomical Society},
	volume       = 156,
	number       = 5,
	pages        = 196,
	doi          = {10.3847/1538-3881/aade95},
	url          = {https://dx.doi.org/10.3847/1538-3881/aade95},
}

@article{Balmer_2025,
	title        = {JWST-TST High Contrast: Living on the Wedge, or, NIRCam Bar Coronagraphy Reveals CO2 in the HR 8799 and 51 Eri Exoplanets' Atmospheres},
	author       = {Balmer, William O. and Kammerer, Jens and Pueyo, Laurent and Perrin, Marshall D. and Girard, Julien H. and Leisenring, Jarron M. and Lawson, Kellen and Dennen, Henry and van der Marel, Roeland P. and Beichman, Charles A. and Bryden, Geoffrey and Llop-Sayson, Jorge and Valenti, Jeff A. and Lothringer, Joshua D. and Lewis, Nikole K. and M\^{a}lin, Mathilde and Rebollido, Isabel and Rickman, Emily and Hoch, Kielan K. W. and Soummer, R\'{e}mi and Clampin, Mark and Mountain, C. Matt},
	year         = 2025,
	month        = mar,
	journal      = {\aj},
	publisher    = {American Astronomical Society},
	volume       = 169,
	number       = 4,
	pages        = 209,
	doi          = {10.3847/1538-3881/adb1c6},
	issn         = {1538-3881},
	url          = {http://dx.doi.org/10.3847/1538-3881/adb1c6},
}

@ARTICLE{Greenbaum2019,
       author = {{Greenbaum}, Alexandra Z. and {Cheetham}, Anthony and {Sivaramakrishnan}, Anand and {Rantakyr{\"o}}, Fredrik T. and {Duch{\^e}ne}, Gaspard and {Tuthill}, Peter and {De Rosa}, Robert J. and {Oppenheimer}, Rebecca and {Macintosh}, Bruce and {Ammons}, S. Mark and {Bailey}, Vanessa P. and {Barman}, Travis and {Bulger}, Joanna and {Cardwell}, Andrew and {Chilcote}, Jeffrey and {Cotten}, Tara and {Doyon}, Rene and {Fitzgerald}, Michael P. and {Follette}, Katherine B. and {Gerard}, Benjamin L. and {Goodsell}, Stephen J. and {Graham}, James R. and {Hibon}, Pascale and {Hung}, Li-Wei and {Ingraham}, Patrick and {Kalas}, Paul and {Konopacky}, Quinn and {Larkin}, James E. and {Maire}, J{\'e}r{\^o}me and {Marchis}, Franck and {Marley}, Mark S. and {Marois}, Christian and {Metchev}, Stanimir and {Millar-Blanchaer}, Maxwell A. and {Morzinski}, Katie M. and {Nielsen}, Eric L. and {Palmer}, David and {Patience}, Jennifer and {Perrin}, Marshall and {Poyneer}, Lisa and {Pueyo}, Laurent and {Rajan}, Abhijith and {Rameau}, Julien and {Sadakuni}, Naru and {Savransky}, Dmitry and {Schneider}, Adam C. and {Song}, Inseok and {Soummer}, Remi and {Thomas}, Sandrine and {Wallace}, J. Kent and {Wang}, Jason J. and {Ward-Duong}, Kimberly and {Wiktorowicz}, Sloane and {Wolff}, Schuyler},
        title = "{Performance of the Gemini Planet Imager Non-redundant Mask and Spectroscopy of Two Close-separation Binaries: HR 2690 and HD 142527}",
      journal = {\aj},
         year = 2019,
        month = jun,
       volume = {157},
       number = {6},
          eid = {249},
        pages = {249},
          doi = {10.3847/1538-3881/ab17db},
archivePrefix = {arXiv},
       eprint = {1904.09006},
 primaryClass = {astro-ph.IM},
       adsurl = {https://ui.adsabs.harvard.edu/abs/2019AJ....157..249G}
}

@ARTICLE{Ruder2016,
       author = {{Ruder}, Sebastian},
        title = "{An overview of gradient descent optimization algorithms}",
      journal = {arXiv e-prints},
         year = 2016,
        month = sep,
          eid = {arXiv:1609.04747},
        pages = {arXiv:1609.04747},
          doi = {10.48550/arXiv.1609.04747},
archivePrefix = {arXiv},
       eprint = {1609.04747},
 primaryClass = {cs.LG},
       adsurl = {https://ui.adsabs.harvard.edu/abs/2016arXiv160904747R}
}

@ARTICLE{Betancourt2017,
       author = {{Betancourt}, Michael},
        title = "{A Conceptual Introduction to Hamiltonian Monte Carlo}",
      journal = {arXiv e-prints},
         year = 2017,
        month = jan,
          eid = {arXiv:1701.02434},
        pages = {arXiv:1701.02434},
          doi = {10.48550/arXiv.1701.02434},
archivePrefix = {arXiv},
       eprint = {1701.02434},
 primaryClass = {stat.ME},
       adsurl = {https://ui.adsabs.harvard.edu/abs/2017arXiv170102434B}
}

@ARTICLE{Metropolis1953,
       author = {{Metropolis}, Nicholas and {Rosenbluth}, Arianna W. and {Rosenbluth}, Marshall N. and {Teller}, Augusta H. and {Teller}, Edward},
        title = "{Equation of State Calculations by Fast Computing Machines}",
      journal = {\jcp},
         year = 1953,
        month = jun,
       volume = {21},
       number = {6},
        pages = {1087-1092},
          doi = {10.1063/1.1699114},
       adsurl = {https://ui.adsabs.harvard.edu/abs/1953JChPh..21.1087M}
}

@article{numpyro_1,
  title = {Composable {{Effects}} for {{Flexible}} and {{Accelerated Probabilistic Programming}} in {{NumPyro}}},
  author = {Phan, Du and Pradhan, Neeraj and Jankowiak, Martin},
  year = 2019,
  month = dec,
  publisher = {arXiv},
  doi = {10.48550/arXiv.1912.11554},
  urldate = {2026-09-03}
}

@article{numpyro_2,
	title        = {Pyro: Deep Universal Probabilistic Programming},
	author       = {Eli Bingham and Jonathan P. Chen and Martin Jankowiak and Fritz Obermeyer and Neeraj Pradhan and Theofanis Karaletsos and Rohit Singh and Paul A. Szerlip and Paul Horsfall and Noah D. Goodman},
	year         = 2019,
	journal      = {J. Mach. Learn. Res.},
	volume       = 20,
	pages        = {28:1--28:6},
	url          = {http://jmlr.org/papers/v20/18-403.html},
}

@article{Cheetham2012,
	title        = {Fizeau interferometric cophasing of segmented mirrors},
	author       = {{Cheetham}, Anthony C. and {Tuthill}, Peter G. and {Sivaramakrishnan}, Anand and {Lloyd}, James P.},
	year         = 2012,
	month        = dec,
	journal      = {Optics Express},
	volume       = 20,
	number       = 28,
	pages        = 29457,
	doi          = {10.1364/OE.20.029457},
	adsurl       = {https://ui.adsabs.harvard.edu/abs/2012OExpr..2029457C},
}

@article{Benner2015,
	title        = {A Survey of Projection-Based Model Reduction Methods for Parametric Dynamical Systems},
	author       = {Benner, Peter and Gugercin, Serkan and Willcox, Karen},
	year         = 2015,
	journal      = {SIAM Review},
	volume       = 57,
	number       = 4,
	pages        = {483--531},
	doi          = {10.1137/130932715},
	url          = {https://doi.org/10.1137/130932715},
	eprint       = {https://doi.org/10.1137/130932715},
}

@article{charlesImageReconstructionJWST2026c,
  title = {Image Reconstruction with the {{JWST}} Interferometer},
  author = {Charles, Max and Desdoigts, Louis and Pope, Benjamin and Tuthill, Peter and Blakely, Dori and Johnstone, Doug and Ray, Shrishmoy and Ford, K. E. Saavik and McKernan, Barry and Sivaramakrishnan, Anand},
  year = 2026,
  month = jan,
  journal = {Publications of the Astronomical Society of Australia},
  volume = {43},
  pages = {e048},
  issn = {1323-3580, 1448-6083},
  doi = {10.1017/pasa.2026.10179},
  urldate = {2026-09-03},
  langid = {english}
}

@article{desdoigtsAmigoDatadrivenCalibration2026,
  title = {Amigo: {{A}} Data-Driven Calibration of the {{JWST}} Interferometer},
  shorttitle = {Amigo},
  author = {Desdoigts, Louis and Pope, Benjamin and Charles, Max and Tuthill, Peter and Blakely, Dori and Johnstone, Douglas and Ray, Shrishmoy and Sivaramakrishnan, Anand and Volk, Kevin and Kammerer, Jens and Thatte, Deepashri and Cooper, Rachel},
  year = 2026,
  month = jan,
  journal = {Publications of the Astronomical Society of Australia},
  volume = {43},
  pages = {e075},
  issn = {1323-3580, 1448-6083},
  doi = {10.1017/pasa.2026.10194},
  urldate = {2026-09-03},
  langid = {english}
}

@ARTICLE{Ray2023b,
  title = {The {{JWST Early Release Science Program}} for {{Direct Observations}} of {{Exoplanetary Systems}}. {{III}}. {{Aperture Masking Interferometric Observations}} of the {{Star HIP}} 65426 at 3.8 {$\mu$}m},
  author = {Ray, Shrishmoy and Sallum, Steph and Hinkley, Sasha and Sivaramkrishnan, Anand and Cooper, Rachel and Kammerer, Jens and Greebaum, Alexandra Z. and Thatte, Deeparshi and Stolker, Tomas and Lazzoni, Cecilia and Tokovinin, Andrei and {de Furio}, Matthew and Factor, Samuel and Meyer, Michael and Stone, Jordan M. and Carter, Aarynn and Biller, Beth and Skemer, Andrew and Su{\'a}rez, Genaro and Leisenring, Jarron M. and Perrin, Marshall D. and Kraus, Adam L. and Absil, Olivier and Balmer, William O. and Boccaletti, Anthony and Bonavita, Mariangela and Bonnefoy, Mickael and Booth, Mark and Bowler, Brendan P. and Briesemeister, Zackery W. and Bryan, Marta L. and Calissendorff, Per and Cantalloube, Faustine and Chauvin, Gael and Chen, Christine H. and Choquet, Elodie and Christiaens, Valentin and Cugno, Gabriele and Currie, Thayne and Danielski, Camilla and Dupuy, Trent J. and Faherty, Jacqueline K. and Fitzgerald, Michael P. and Fortney, Jonathan J. and Franson, Kyle and Girard, Julien H. and Grady, Carol A. and Gonzales, Eileen C. and Henning, Thomas and Hines, Dean C. and Hoch, Kielan K. W. and Hood, Callie E. and Howe, Alex R. and Janson, Markus and Kalas, Paul and Kennedy, Grant M. and Kenworthy, Matthew A. and Kervella, Pierre and Kuzuhara, Masayuki and Lagrange, Anne-Marie and Lagage, Pierre-Olivier and Lawson, Kellen and Lew, Ben W. P. and Liu, Michael C. and Liu, Pengyu and {Llop-Sayson}, Jorge and Lloyd, James P. and Macintosh, Bruce and Marino, Sebastian and Marley, Mark S. and Marois, Christian and Martinez, Raquel A. and Matthews, Brenda C. and Matthews, Elisabeth C. and Mawet, Dimitri and Mazoyer, Johan and McElwain, Michael W. and Metchev, Stanimir and Meyer, Michael R. and Miles, Brittany E. and {Millar-Blanchaer}, Maxwell A. and Molliere, Paul and Moran, Sarah E. and Morley, Caroline V. and Mukherjee, Sagnick and {Palma-Bifani}, Paulina and Pantin, Eric and Patapis, Polychronis and Petrus, Simon and Pueyo, Laurent and Quanz, Sascha P. and Quirrenbach, Andreas and Rebollido, Isabel and Adams Redai, Jea and Ren, Bin B. and Rickman, Emily and Samland, Matthias and Schlieder, Joshua E. and Schneider, Glenn and Stapelfeldt, Karl R. and Tamura, Motohide and Tan, Xianyu and Uyama, Taichi and Vigan, Arthur and Vasist, Malavika and Vos, Johanna M. and Wagner, Kevin and Wang, Jason J. and {Ward-Duong}, Kimberly and Whiteford, Niall and Wolff, Schuyler G. and Worthen, Kadin and Wyatt, Mark C. and Ygouf, Marie and Zhang, Xi and Zhang, Keming and Zhang, Zhoujian and Zhou, Yifan and Zurlo, Alice and Sargent, B. A. and Theissen, Christopher A. and Manjavacas, Elena and Lueber, Anna and Kitzmann, Daniel and Sutlieff, Ben J. and Betti, Sarah K.},
  year = 2025,
  month = apr,
  journal = {The Astrophysical Journal Letters},
  volume = {983},
  number = {1},
  pages = {L25},
  publisher = {The American Astronomical Society},
  issn = {2041-8205},
  doi = {10.3847/2041-8213/adaeb7},
  urldate = {2026-09-03},
  langid = {english}
}

@ARTICLE{Sallum2023,
  title = {The {{JWST Early Release Science Program}} for {{Direct Observations}} of {{Exoplanetary Systems}}. {{IV}}. {{NIRISS Aperture Masking Interferometry Performance}} and {{Lessons Learned}}},
  author = {Sallum, Steph and Ray, Shrishmoy and Kammerer, Jens and Sivaramakrishnan, Anand and Cooper, Rachel and Greebaum, Alexandra Z. and Thatte, Deepashri and De Furio, Matthew and Factor, Samuel M. and Meyer, Michael R. and Stone, Jordan M. and Carter, Aarynn and Biller, Beth and Hinkley, Sasha and Skemer, Andrew and Su{\'a}rez, Genaro and Leisenring, Jarron M. and Perrin, Marshall D. and Kraus, Adam L. and Absil, Olivier and Balmer, William O. and Betti, Sarah K. and Boccaletti, Anthony and Bonavita, Mariangela and Bonnefoy, Mickael and Booth, Mark and Bowler, Brendan P. and Briesemeister, Zackery W. and Bryan, Marta L. and Calissendorff, Per and Cantalloube, Faustine and Chauvin, Gael and Chen, Christine H. and Choquet, Elodie and Christiaens, Valentin and Cugno, Gabriele and Currie, Thayne and Danielski, Camilla and Dupuy, Trent J. and Faherty, Jacqueline K. and Fitzgerald, Michael P. and Fortney, Jonathan J. and Franson, Kyle and Girard, Julien H. and Grady, Carol A. and Gonzales, Eileen C. and Henning, Thomas and Hines, Dean C. and Hoch, Kielan K. W. and Hood, Callie E. and Howe, Alex R. and Janson, Markus and Kalas, Paul and Kennedy, Grant M. and Kenworthy, Matthew A. and Kervella, Pierre and Kitzmann, Daniel and Kuzuhara, Masayuki and Lagrange, Anne-Marie and Lagage, Pierre-Olivier and Lawson, Kellen and Lazzoni, Cecilia and Lew, Ben W. P. and Liu, Michael C. and Liu, Pengyu and {Llop-Sayson}, Jorge and Lloyd, James P. and Lueber, Anna and Macintosh, Bruce and Manjavacas, Elena and Marino, Sebastian and Marley, Mark S. and Marois, Christian and Martinez, Raquel A. and Matthews, Brenda C. and Matthews, Elisabeth C. and Mawet, Dimitri and Mazoyer, Johan and McElwain, Michael W. and Metchev, Stanimir and Miles, Brittany E. and {Millar-Blanchaer}, Maxwell A. and Molliere, Paul and Moran, Sarah E. and Morley, Caroline V. and Mukherjee, Sagnick and {Palma-Bifani}, Paulina and Pantin, Eric and Patapis, Polychronis and Petrus, Simon and Pueyo, Laurent and Quanz, Sascha P. and Quirrenbach, Andreas and Rebollido, Isabel and Redai, Jea Adams and Ren, Bin B. and Rickman, Emily and Samland, Matthias and Sargent, B. A. and Schlieder, Joshua E. and Schneider, Glenn and Stapelfeldt, Karl R. and Sutlieff, Ben J. and Tamura, Motohide and Tan, Xianyu and Theissen, Christopher A. and Uyama, Taichi and Vigan, Arthur and Vasist, Malavika and Vos, Johanna M. and Wagner, Kevin and Wang, Jason J. and {Ward-Duong}, Kimberly and Whiteford, Niall and Wolff, Schuyler G. and Worthen, Kadin and Wyatt, Mark C. and Ygouf, Marie and Zhang, Xi and Zhang, Keming and Zhang, Zhoujian and Zhou, Yifan and Zurlo, Alice},
  year = 2024,
  month = feb,
  journal = {The Astrophysical Journal Letters},
  volume = {963},
  number = {1},
  pages = {L2},
  publisher = {The American Astronomical Society},
  issn = {2041-8205},
  doi = {10.3847/2041-8213/ad21fb},
  urldate = {2026-09-03},
  langid = {english}
}

@article{greenbaumIMAGEPLANEALGORITHMJWSTS2014,
  title = {{{AN IMAGE-PLANE ALGORITHM FOR JWST}}'{{S NON-REDUNDANT APERTURE MASK DATA}}},
  author = {Greenbaum, Alexandra Z. and Pueyo, Laurent and Sivaramakrishnan, Anand and Lacour, Sylvestre},
  year = 2014,
  month = dec,
  journal = {The Astrophysical Journal},
  volume = {798},
  number = {2},
  pages = {68},
  publisher = {The American Astronomical Society},
  issn = {0004-637X},
  doi = {10.1088/0004-637X/798/2/68},
  urldate = {2026-09-07},
  langid = {english}
}

@ARTICLE{Octofitter,
       author = {{Thompson}, William and {Lawrence}, Jensen and {Blakely}, Dori and {Marois}, Christian and {Wang}, Jason and {Giordano}, Mos{\'e} and {Brandt}, Timothy and {Johnstone}, Doug and {Ruffio}, Jean-Baptiste and {Ammons}, S. Mark and {Crotts}, Katie A. and {Do {\'O}}, Clarissa R. and {Gonzales}, Eileen C. and {Rice}, Malena},
        title = "{Octofitter: Fast, Flexible, and Accurate Orbit Modeling to Detect Exoplanets}",
      journal = {\aj},
         year = 2023,
        month = oct,
       volume = {166},
       number = {4},
          eid = {164},
        pages = {164},
          doi = {10.3847/1538-3881/acf5cc},
archivePrefix = {arXiv},
       eprint = {2402.01971},
 primaryClass = {astro-ph.EP},
       adsurl = {https://ui.adsabs.harvard.edu/abs/2023AJ....166..164T}
}

@article{Amari1998,
  title = {Natural Gradient Works Efficiently in Learning},
  volume = {10},
  ISSN = {1530-888X},
  url = {http://dx.doi.org/10.1162/089976698300017746},
  DOI = {10.1162/089976698300017746},
  number = {2},
  journal = {Neural Computation},
  publisher = {MIT Press},
  author = {Amari,  Shun-ichi},
  year = {1998},
  month = feb,
  pages = {251–276}
}

@article{gozdziewskiMultipleMeanMotion2014a,
  title = {Multiple Mean Motion Resonances in the {{HR}} 8799 Planetary System},
  author = {Go{\'z}dziewski, Krzysztof and Migaszewski, Cezary},
  year = 2014,
  month = jun,
  journal = {Monthly Notices of the Royal Astronomical Society},
  volume = {440},
  pages = {3140--3171},
  publisher = {OUP},
  issn = {0035-8711},
  doi = {10.1093/mnras/stu455},
  urldate = {2026-04-03}
}

@article{thompsonDeepOrbitalSearch2022,
  title = {Deep {{Orbital Search}} for {{Additional Planets}} in the {{HR}} 8799 {{System}}},
  author = {Thompson, William and Marois, Christian and Do {\'O}, Clarissa R. and Konopacky, Quinn and Ruffio, Jean-Baptiste and Wang, Jason and Skemer, Andy J. and De Rosa, Robert J. and Macintosh, Bruce},
  year = 2022,
  month = dec,
  journal = {The Astronomical Journal},
  volume = {165},
  number = {1},
  pages = {29},
  publisher = {The American Astronomical Society},
  issn = {1538-3881},
  doi = {10.3847/1538-3881/aca1af},
  urldate = {2026-03-13},
  langid = {english}
}

@article{faramazDetailedCharacterizationHR2021,
  title = {A {{Detailed Characterization}} of {{HR}} 8799's {{Debris Disk}} with {{ALMA}} in {{Band}} 7},
  author = {Faramaz, Virginie and Marino, Sebastian and Booth, Mark and Matr{\`a}, Luca and Mamajek, Eric E. and Bryden, Geoffrey and Stapelfeldt, Karl R. and Casassus, Simon and Cuadra, Jorge and Hales, Antonio S. and Zurlo, Alice},
  year = 2021,
  month = may,
  journal = {The Astronomical Journal},
  volume = {161},
  number = {6},
  pages = {271},
  publisher = {The American Astronomical Society},
  issn = {1538-3881},
  doi = {10.3847/1538-3881/abf4e0},
  urldate = {2026-04-15},
  langid = {english}
}

@article{zurloOrbitalDynamicalAnalysis2022,
  title = {Orbital and Dynamical Analysis of the System around {{HR}} 8799. {{New}} Astrometric Epochs from {{VLT}}/{{SPHERE}} and {{LBT}}/{{LUCI}}},
  author = {Zurlo, A. and Go{\'z}dziewski, K. and Lazzoni, C. and Mesa, D. and Nogueira, P. and Desidera, S. and Gratton, R. and Marzari, F. and Langlois, M. and Pinna, E. and Chauvin, G. and Delorme, P. and Girard, J. H. and Hagelberg, J. and Henning, {\relax Th}. and Janson, M. and Rickman, E. and Kervella, P. and Avenhaus, H. and Bhowmik, T. and Biller, B. and Boccaletti, A. and Bonaglia, M. and Bonavita, M. and Bonnefoy, M. and Cantalloube, F. and Cheetham, A. and Claudi, R. and D'Orazi, V. and Feldt, M. and Galicher, R. and Ghose, E. and Lagrange, A.-M. and {le Coroller}, H. and Ligi, R. and Kasper, M. and Maire, A.-L. and Medard, F. and Meyer, M. and Peretti, S. and Perrot, C. and Puglisi, A. T. and Rossi, F. and Rothberg, B. and Schmidt, T. and Sissa, E. and Vigan, A. and Wahhaj, Z.},
  year = 2022,
  month = oct,
  journal = {Astronomy and Astrophysics},
  volume = {666},
  pages = {A133},
  publisher = {EDP},
  issn = {0004-6361},
  doi = {10.1051/0004-6361/202243862},
  urldate = {2026-04-15}
}

@article{maroisImagesFourthPlanet2010a,
  title = {Images of a Fourth Planet Orbiting {{HR}} 8799},
  author = {Marois, Christian and Zuckerman, B. and Konopacky, Quinn M. and Macintosh, Bruce and Barman, Travis},
  year = 2010,
  month = dec,
  journal = {Nature},
  volume = {468},
  pages = {1080--1083},
  issn = {0028-0836},
  doi = {10.1038/nature09684},
  urldate = {2026-04-15}
}

@article{sivaramakrishnanInfraredImagerSlitless2023,
  title = {The {{Near Infrared Imager}} and {{Slitless Spectrograph}} for the {{James Webb Space Telescope}}. {{IV}}. {{Aperture Masking Interferometry}}},
  author = {Sivaramakrishnan, Anand and Tuthill, Peter and Lloyd, James P. and Greenbaum, Alexandra Z. and Thatte, Deepashri and Cooper, Rachel A. and Vandal, Thomas and Kammerer, Jens and {Sanchez-Bermudez}, Joel and Pope, Benjamin J. S. and Blakely, Dori and Albert, Lo{\"i}c and Cook, Neil J. and Johnstone, Doug and Martel, Andr{\'e} R. and Volk, Kevin and Soulain, Anthony and Artigau, {\'E}tienne and Lafreni{\`e}re, David and Willott, Chris J. and Parmentier, S{\'e}bastien and Ford, K. E. Saavik and McKernan, Barry and Vila, M. Bego{\~n}a and Rowlands, Neil and Doyon, Ren{\'e} and Beaulieu, Mathilde and Desdoigts, Louis and Fullerton, Alexander W. and De Furio, Matthew and Goudfrooij, Paul and Holfeltz, Sherie T. and LaMassa, Stephanie and Maszkiewicz, Michael and Meyer, Michael R. and Perrin, Marshall D. and Pueyo, Laurent and Sahlmann, Johannes and Sohn, Sangmo Tony and Teixeira, Paula S. and Zheng, Sheng-hai},
  year = 2023,
  month = feb,
  journal = {Publications of the Astronomical Society of the Pacific},
  volume = {135},
  number = {1043},
  pages = {015003},
  publisher = {The Astronomical Society of the Pacific},
  issn = {1538-3873},
  doi = {10.1088/1538-3873/acaebd},
  urldate = {2026-04-15},
  langid = {english}
}

@article{nasedkinFourofakindComprehensiveAtmospheric2024a,
  title = {Four-of-a-Kind? {{Comprehensive}} Atmospheric Characterisation of the {{HR}} 8799 Planets with {{VLTI}}/{{GRAVITY}}},
  shorttitle = {Four-of-a-Kind?},
  author = {Nasedkin, E. and Molli{\`e}re, P. and Lacour, S. and Nowak, M. and Kreidberg, L. and Stolker, T. and Wang, J. J. and Balmer, W. O. and Kammerer, J. and Shangguan, J. and Abuter, R. and Amorim, A. and {Asensio-Torres}, R. and Benisty, M. and Berger, J.-P. and Beust, H. and Blunt, S. and Boccaletti, A. and Bonnefoy, M. and Bonnet, H. and Bordoni, M. S. and Bourdarot, G. and Brandner, W. and Cantalloube, F. and Caselli, P. and Charnay, B. and Chauvin, G. and Chavez, A. and Choquet, E. and Christiaens, V. and Cl{\'e}net, Y. and du Foresto, V. Coud{\'e} and Cridland, A. and Davies, R. and Dembet, R. and Dexter, J. and Drescher, A. and Duvert, G. and Eckart, A. and Eisenhauer, F. and Schreiber, N. M. F{\"o}rster and Garcia, P. and Lopez, R. Garcia and Gendron, E. and Genzel, R. and Gillessen, S. and Girard, J. H. and Grant, S. and Haubois, X. and Hei{\ss}el, G. and Henning, Th and Hinkley, S. and Hippler, S. and Houll{\'e}, M. and Hubert, Z. and Jocou, L. and Keppler, M. and Kervella, P. and Kurtovic, N. T. and Lagrange, A.-M. and Lapeyr{\`e}re, V. and Bouquin, J.-B. Le and Lutz, D. and Maire, A.-L. and Mang, F. and Marleau, G.-D. and M{\'e}rand, A. and Monnier, J. D. and Mordasini, C. and Ott, T. and Otten, G. P. P. L. and Paladini, C. and Paumard, T. and Perraut, K. and Perrin, G. and Pfuhl, O. and Pourr{\'e}, N. and Pueyo, L. and Ribeiro, D. C. and Rickman, E. and Ruffio, J. B. and Rustamkulov, Z. and Shimizu, T. and Sing, D. and Stadler, J. and Straub, O. and Straubmeier, C. and Sturm, E. and Tacconi, L. J. and van Dishoeck, E. F. and Vigan, A. and Vincent, F. and von Fellenberg, S. D. and Widmann, F. and Winterhalder, T. O. and Woillez, J. and Yazici, {\c S}},
  year = 2024,
  month = jul,
  journal = {Astronomy \& Astrophysics},
  volume = {687},
  pages = {A298},
  publisher = {EDP Sciences},
  issn = {0004-6361, 1432-0746},
  doi = {10.1051/0004-6361/202449328},
  urldate = {2026-04-15},
  copyright = {\copyright{} The Authors 2024},
  langid = {english}
}

@article{irelandPhaseErrorsDiffractionlimited2013,
  title = {Phase Errors in Diffraction-Limited Imaging: Contrast Limits for Sparse Aperture Masking},
  shorttitle = {Phase Errors in Diffraction-Limited Imaging},
  author = {Ireland, M. J.},
  year = 2013,
  month = aug,
  journal = {Monthly Notices of the Royal Astronomical Society},
  volume = {433},
  number = {2},
  pages = {1718--1728},
  issn = {0035-8711},
  doi = {10.1093/mnras/stt859},
  urldate = {2026-04-19}
}

@article{xuanCompositionsHR87992026,
  title = {The {{Compositions}} of the {{HR}} 8799 {{Planets Reflect Accretion}} of {{Both Solids}} and {{Metal-enriched Gas}}},
  author = {Xuan, Jerry W. and Ruffio, Jean-Baptiste and Chachan, Yayaati and Ohno, Kazumasa and Kesseli, Aurora and {Murray-Clay}, Ruth and Lee, Eve J. and Moses, Julianne I. and Balmer, William O. and Baburaj, Aneesh and Blake, Geoffrey A. and Johnstone, Doug and Zhang, Yapeng and Knutson, Heather A. and Mawet, Dimitri and Beichman, Charles and Hodapp, Klaus and Perrin, Marshall D. and Konopacky, Quinn and Meyer, Michael and Bryden, Geoffrey and Greene, Thomas P. and Leisenring, Jarron and Ygouf, Marie and Benneke, Bj{\"o}rn and Inglis, Julie and Wallack, Nicole L.},
  year = 2026,
  month = mar,
  journal = {The Astrophysical Journal},
  volume = {1000},
  number = {1},
  pages = {27},
  publisher = {The American Astronomical Society},
  issn = {0004-637X},
  doi = {10.3847/1538-4357/ae448f},
  urldate = {2026-04-15},
  langid = {english}
}

@article{sepulvedaDynamicalMassExoplanet2022,
  title = {Dynamical {{Mass}} of the {{Exoplanet Host Star HR}} 8799},
  author = {Sepulveda, Aldo G. and Bowler, Brendan P.},
  year = 2022,
  month = jan,
  journal = {The Astronomical Journal},
  volume = {163},
  number = {2},
  pages = {52},
  publisher = {The American Astronomical Society},
  issn = {1538-3881},
  doi = {10.3847/1538-3881/ac3bb5},
  urldate = {2026-04-23},
  langid = {english}
}

@article{gozdziewskiExactGeneralizedLaplace2020a,
  title = {An {{Exact}}, {{Generalized Laplace Resonance}} in the {{HR}} 8799 {{Planetary System}}},
  author = {Go{\'z}dziewski, Krzysztof and Migaszewski, Cezary},
  year = 2020,
  month = oct,
  journal = {The Astrophysical Journal Letters},
  volume = {902},
  number = {2},
  pages = {L40},
  publisher = {The American Astronomical Society},
  issn = {2041-8205},
  doi = {10.3847/2041-8213/abb881},
  urldate = {2026-08-01},
  langid = {english}
}

@article{bainesCHARAARRAYANGULAR2012,
  title = {{{THE CHARA ARRAY ANGULAR DIAMETER OF HR}} 8799 {{FAVORS PLANETARY MASSES FOR ITS IMAGED COMPANIONS}}},
  author = {Baines, Ellyn K. and White, Russel J. and Huber, Daniel and Jones, Jeremy and Boyajian, Tabetha and McAlister, Harold A. and {ten Brummelaar}, Theo A. and Turner, Nils H. and Sturmann, Judit and Sturmann, Laszlo and Goldfinger, P. J. and Farrington, Christopher D. and Riedel, Adric R. and Ireland, Michael and {von Braun}, Kaspar and Ridgway, Stephen T.},
  year = 2012,
  month = nov,
  journal = {The Astrophysical Journal},
  volume = {761},
  number = {1},
  pages = {57},
  publisher = {The American Astronomical Society},
  issn = {0004-637X},
  doi = {10.1088/0004-637X/761/1/57},
  urldate = {2026-08-10},
  langid = {english}
}

@article{surjanovicPigeonsjlDistributedSampling2025,
  title = {Pigeons.Jl: {{Distributed}} Sampling from Intractable Distributions},
  shorttitle = {Pigeons.Jl},
  author = {Surjanovic, Nikola and {Biron-Lattes}, Miguel and Tiede, Paul and Syed, Saifuddin and Campbell, Trevor and {Bouchard-C{\textbackslash}}, Alexandre},
  year = 2025,
  month = mar,
  journal = {Proceedings of the JuliaCon Conferences},
  volume = {7},
  number = {69},
  pages = {139},
  issn = {2642-4029},
  doi = {10.21105/jcon.00139},
  urldate = {2026-09-03},
  langid = {english}
}

@article{husserNewExtensiveLibrary2013a,
  title = {A New Extensive Library of {{PHOENIX}} Stellar Atmospheres and Synthetic Spectra},
  author = {Husser, T.-O. and Berg, S. Wende-von and Dreizler, S. and Homeier, D. and Reiners, A. and Barman, T. and Hauschildt, P. H.},
  year = 2013,
  month = may,
  journal = {Astronomy \& Astrophysics},
  volume = {553},
  pages = {A6},
  publisher = {EDP Sciences},
  issn = {0004-6361, 1432-0746},
  doi = {10.1051/0004-6361/201219058},
  urldate = {2026-08-14},
  copyright = {\copyright{} ESO, 2013},
  langid = {english}
}

@article{morleySonoraSubstellarAtmosphere2024,
  title = {The {{Sonora Substellar Atmosphere Models}}. {{III}}. {{Diamondback}}: {{Atmospheric Properties}}, {{Spectra}}, and {{Evolution}} for {{Warm Cloudy Substellar Objects}}},
  shorttitle = {The {{Sonora Substellar Atmosphere Models}}. {{III}}. {{Diamondback}}},
  author = {Morley, Caroline V. and Mukherjee, Sagnick and Marley, Mark S. and Fortney, Jonathan J. and Visscher, Channon and Lupu, Roxana and {Gharib-Nezhad}, Ehsan and Thorngren, Daniel and Freedman, Richard and Batalha, Natasha},
  year = 2024,
  month = oct,
  journal = {The Astrophysical Journal},
  volume = {975},
  number = {1},
  pages = {59},
  publisher = {The American Astronomical Society},
  issn = {0004-637X},
  doi = {10.3847/1538-4357/ad71d5},
  urldate = {2026-08-14},
  langid = {english}
}

@article{ruffioJupiterlikeUniformMetal2026,
  title = {Jupiter-like Uniform Metal Enrichment in a System of Multiple Giant Exoplanets},
  author = {Ruffio, Jean-Baptiste and Xuan, Jerry W. and Chachan, Yayaati and Kesseli, Aurora and Lee, Eve J. and Beichman, Charles and Hodapp, Klaus and Balmer, William O. and Konopacky, Quinn and Perrin, Marshall D. and Mawet, Dimitri and Knutson, Heather A. and Bryden, Geoffrey and Greene, Thomas P. and Johnstone, Doug and Leisenring, Jarron and Meyer, Michael and Ygouf, Marie},
  year = 2026,
  month = apr,
  journal = {Nature Astronomy},
  volume = {10},
  number = {4},
  pages = {511--521},
  publisher = {Nature Publishing Group},
  issn = {2397-3366},
  doi = {10.1038/s41550-026-02783-z},
  urldate = {2026-08-17},
  copyright = {2026 The Author(s), under exclusive licence to Springer Nature Limited},
  langid = {english}
}

@article{dooOrbitalEccentricitiesDirectly2023a,
  title = {The {{Orbital Eccentricities}} of {{Directly Imaged Companions Using Observable-based Priors}}: {{Implications}} for {{Population-level Distributions}}},
  shorttitle = {The {{Orbital Eccentricities}} of {{Directly Imaged Companions Using Observable-based Priors}}},
  author = {Do {\'O}, Clarissa R. and O'Neil, Kelly K. and Konopacky, Quinn M. and Do, Tuan and Martinez, Gregory D. and Ruffio, Jean-Baptiste and Ghez, Andrea M.},
  year = 2023,
  month = jul,
  journal = {The Astronomical Journal},
  volume = {166},
  number = {2},
  pages = {48},
  publisher = {The American Astronomical Society},
  issn = {1538-3881},
  doi = {10.3847/1538-3881/acdc9a},
  urldate = {2026-08-18},
  langid = {english}
}

@article{oneilImprovingOrbitEstimates2019,
  title = {Improving {{Orbit Estimates}} for {{Incomplete Orbits}} with a {{New Approach}} to {{Priors}}: With {{Applications}} from {{Black Holes}} to {{Planets}}},
  shorttitle = {Improving {{Orbit Estimates}} for {{Incomplete Orbits}} with a {{New Approach}} to {{Priors}}},
  author = {O'Neil, K. Kosmo and Martinez, G. D. and Hees, A. and Ghez, A. M. and Do, T. and Witzel, G. and Konopacky, Q. and Becklin, E. E. and Chu, D. S. and Lu, J. R. and Matthews, K. and Sakai, S.},
  year = 2019,
  month = jun,
  journal = {The Astronomical Journal},
  volume = {158},
  number = {1},
  pages = {4},
  publisher = {The American Astronomical Society},
  issn = {1538-3881},
  doi = {10.3847/1538-3881/ab1d66},
  urldate = {2026-08-18},
  langid = {english}
}

@article{wahhajSearchFifthPlanet2021,
  title = {A Search for a Fifth Planet around {{HR}} 8799 Using the Star-Hopping {{RDI}} Technique at {{VLT}}/{{SPHERE}}},
  author = {Wahhaj, Z. and Milli, J. and Romero, C. and Cieza, L. and Zurlo, A. and Vigan, A. and Pe{\~n}a, E. and Valdes, G. and Cantalloube, F. and Girard, J. and Pantoja, B.},
  year = 2021,
  month = apr,
  journal = {Astronomy \& Astrophysics},
  volume = {648},
  pages = {A26},
  publisher = {EDP Sciences},
  issn = {0004-6361, 1432-0746},
  doi = {10.1051/0004-6361/202038794},
  urldate = {2026-08-21},
  copyright = {\copyright{} ESO 2021},
  langid = {english}
}

@article{lucasADIjlJuliaPackage2020,
  title = {{{ADI}}.Jl: {{A Julia Package}} for {{High-Contrast Imaging}}},
  shorttitle = {{{ADI}}.Jl},
  author = {Lucas, Miles and Bottom, Michael},
  year = 2020,
  month = dec,
  journal = {Journal of Open Source Software},
  volume = {5},
  number = {56},
  pages = {2843},
  issn = {2475-9066},
  doi = {10.21105/joss.02843},
  urldate = {2026-08-22},
  langid = {english}
}

@article{astropy:2013,
        Adsurl = {https://adsabs.harvard.edu/abs/2013A%26A...558A..33A},
        Archiveprefix = {arXiv},
        Author = {{Astropy Collaboration} and {Robitaille}, T.~P. and {Tollerud}, E.~J. and {Greenfield}, P. and {Droettboom}, M. and {Bray}, E. and {Aldcroft}, T. and {Davis}, M. and {Ginsburg}, A. and {Price-Whelan}, A.~M. and {Kerzendorf}, W.~E. and {Conley}, A. and {Crighton}, N. and {Barbary}, K. and {Muna}, D. and {Ferguson}, H. and {Grollier}, F. and {Parikh}, M.~M. and {Nair}, P.~H. and {Unther}, H.~M. and {Deil}, C. and {Woillez}, J. and {Conseil}, S. and {Kramer}, R. and {Turner}, J.~E.~H. and {Singer}, L. and {Fox}, R. and {Weaver}, B.~A. and {Zabalza}, V. and {Edwards}, Z.~I. and {Azalee Bostroem}, K. and {Burke}, D.~J. and {Casey}, A.~R. and {Crawford}, S.~M. and {Dencheva}, N. and {Ely}, J. and {Jenness}, T. and {Labrie}, K. and {Lim}, P.~L. and {Pierfederici}, F. and {Pontzen}, A. and {Ptak}, A. and {Refsdal}, B. and {Servillat}, M. and {Streicher}, O.},
        Doi = {10.1051/0004-6361/201322068},
        Eid = {A33},
        Eprint = {1307.6212},
        Journal = {\aap},
        Month = oct,
        Pages = {A33},
        Primaryclass = {astro-ph.IM},
        Title = {{Astropy: A community Python package for astronomy}},
        Volume = 558,
        Year = 2013}

@ARTICLE{astropy:2018,
               author = {{Astropy Collaboration} and {Price-Whelan}, A.~M. and
                 {Sip{\H{o}}cz}, B.~M. and {G{\"u}nther}, H.~M. and {Lim}, P.~L. and
                 {Crawford}, S.~M. and {Conseil}, S. and {Shupe}, D.~L. and
                 {Craig}, M.~W. and {Dencheva}, N. and {Ginsburg}, A. and {Vand
                erPlas}, J.~T. and {Bradley}, L.~D. and {P{\'e}rez-Su{\'a}rez}, D. and
                 {de Val-Borro}, M. and {Aldcroft}, T.~L. and {Cruz}, K.~L. and
                 {Robitaille}, T.~P. and {Tollerud}, E.~J. and {Ardelean}, C. and
                 {Babej}, T. and {Bach}, Y.~P. and {Bachetti}, M. and {Bakanov}, A.~V. and
                 {Bamford}, S.~P. and {Barentsen}, G. and {Barmby}, P. and
                 {Baumbach}, A. and {Berry}, K.~L. and {Biscani}, F. and {Boquien}, M. and
                 {Bostroem}, K.~A. and {Bouma}, L.~G. and {Brammer}, G.~B. and
                 {Bray}, E.~M. and {Breytenbach}, H. and {Buddelmeijer}, H. and
                 {Burke}, D.~J. and {Calderone}, G. and {Cano Rodr{\'\i}guez}, J.~L. and
                 {Cara}, M. and {Cardoso}, J.~V.~M. and {Cheedella}, S. and {Copin}, Y. and
                 {Corrales}, L. and {Crichton}, D. and {D'Avella}, D. and {Deil}, C. and
                 {Depagne}, {\'E}. and {Dietrich}, J.~P. and {Donath}, A. and
                 {Droettboom}, M. and {Earl}, N. and {Erben}, T. and {Fabbro}, S. and
                 {Ferreira}, L.~A. and {Finethy}, T. and {Fox}, R.~T. and
                 {Garrison}, L.~H. and {Gibbons}, S.~L.~J. and {Goldstein}, D.~A. and
                 {Gommers}, R. and {Greco}, J.~P. and {Greenfield}, P. and
                 {Groener}, A.~M. and {Grollier}, F. and {Hagen}, A. and {Hirst}, P. and
                 {Homeier}, D. and {Horton}, A.~J. and {Hosseinzadeh}, G. and {Hu}, L. and
                 {Hunkeler}, J.~S. and {Ivezi{\'c}}, {\v{Z}}. and {Jain}, A. and
                 {Jenness}, T. and {Kanarek}, G. and {Kendrew}, S. and {Kern}, N.~S. and
                 {Kerzendorf}, W.~E. and {Khvalko}, A. and {King}, J. and {Kirkby}, D. and
                 {Kulkarni}, A.~M. and {Kumar}, A. and {Lee}, A. and {Lenz}, D. and
                 {Littlefair}, S.~P. and {Ma}, Z. and {Macleod}, D.~M. and
                 {Mastropietro}, M. and {McCully}, C. and {Montagnac}, S. and
                 {Morris}, B.~M. and {Mueller}, M. and {Mumford}, S.~J. and {Muna}, D. and
                 {Murphy}, N.~A. and {Nelson}, S. and {Nguyen}, G.~H. and
                 {Ninan}, J.~P. and {N{\"o}the}, M. and {Ogaz}, S. and {Oh}, S. and
                 {Parejko}, J.~K. and {Parley}, N. and {Pascual}, S. and {Patil}, R. and
                 {Patil}, A.~A. and {Plunkett}, A.~L. and {Prochaska}, J.~X. and
                 {Rastogi}, T. and {Reddy Janga}, V. and {Sabater}, J. and
                 {Sakurikar}, P. and {Seifert}, M. and {Sherbert}, L.~E. and
                 {Sherwood-Taylor}, H. and {Shih}, A.~Y. and {Sick}, J. and
                 {Silbiger}, M.~T. and {Singanamalla}, S. and {Singer}, L.~P. and
                 {Sladen}, P.~H. and {Sooley}, K.~A. and {Sornarajah}, S. and
                 {Streicher}, O. and {Teuben}, P. and {Thomas}, S.~W. and
                 {Tremblay}, G.~R. and {Turner}, J.~E.~H. and {Terr{\'o}n}, V. and
                 {van Kerkwijk}, M.~H. and {de la Vega}, A. and {Watkins}, L.~L. and
                 {Weaver}, B.~A. and {Whitmore}, J.~B. and {Woillez}, J. and
                 {Zabalza}, V. and {Astropy Contributors}},
                title = "{The Astropy Project: Building an Open-science Project and Status of the v2.0 Core Package}",
              journal = {\aj},
                 year = 2018,
                month = sep,
               volume = {156},
               number = {3},
                  eid = {123},
                pages = {123},
                  doi = {10.3847/1538-3881/aabc4f},
        archivePrefix = {arXiv},
               eprint = {1801.02634},
         primaryClass = {astro-ph.IM},
               adsurl = {https://ui.adsabs.harvard.edu/abs/2018AJ....156..123A}
        }

@ARTICLE{astropy:2022,
               author = {{Astropy Collaboration} and {Price-Whelan}, Adrian M. and {Lim}, Pey Lian and {Earl}, Nicholas and {Starkman}, Nathaniel and {Bradley}, Larry and {Shupe}, David L. and {Patil}, Aarya A. and {Corrales}, Lia and {Brasseur}, C.~E. and {N{"o}the}, Maximilian and {Donath}, Axel and {Tollerud}, Erik and {Morris}, Brett M. and {Ginsburg}, Adam and {Vaher}, Eero and {Weaver}, Benjamin A. and {Tocknell}, James and {Jamieson}, William and {van Kerkwijk}, Marten H. and {Robitaille}, Thomas P. and {Merry}, Bruce and {Bachetti}, Matteo and {G{"u}nther}, H. Moritz and {Aldcroft}, Thomas L. and {Alvarado-Montes}, Jaime A. and {Archibald}, Anne M. and {B{'o}di}, Attila and {Bapat}, Shreyas and {Barentsen}, Geert and {Baz{'a}n}, Juanjo and {Biswas}, Manish and {Boquien}, M{'e}d{'e}ric and {Burke}, D.~J. and {Cara}, Daria and {Cara}, Mihai and {Conroy}, Kyle E. and {Conseil}, Simon and {Craig}, Matthew W. and {Cross}, Robert M. and {Cruz}, Kelle L. and {D'Eugenio}, Francesco and {Dencheva}, Nadia and {Devillepoix}, Hadrien A.~R. and {Dietrich}, J{"o}rg P. and {Eigenbrot}, Arthur Davis and {Erben}, Thomas and {Ferreira}, Leonardo and {Foreman-Mackey}, Daniel and {Fox}, Ryan and {Freij}, Nabil and {Garg}, Suyog and {Geda}, Robel and {Glattly}, Lauren and {Gondhalekar}, Yash and {Gordon}, Karl D. and {Grant}, David and {Greenfield}, Perry and {Groener}, Austen M. and {Guest}, Steve and {Gurovich}, Sebastian and {Handberg}, Rasmus and {Hart}, Akeem and {Hatfield-Dodds}, Zac and {Homeier}, Derek and {Hosseinzadeh}, Griffin and {Jenness}, Tim and {Jones}, Craig K. and {Joseph}, Prajwel and {Kalmbach}, J. Bryce and {Karamehmetoglu}, Emir and {Ka{l}uszy{'n}ski}, Miko{l}aj and {Kelley}, Michael S.~P. and {Kern}, Nicholas and {Kerzendorf}, Wolfgang E. and {Koch}, Eric W. and {Kulumani}, Shankar and {Lee}, Antony and {Ly}, Chun and {Ma}, Zhiyuan and {MacBride}, Conor and {Maljaars}, Jakob M. and {Muna}, Demitri and {Murphy}, N.~A. and {Norman}, Henrik and {O'Steen}, Richard and {Oman}, Kyle A. and {Pacifici}, Camilla and {Pascual}, Sergio and {Pascual-Granado}, J. and {Patil}, Rohit R. and {Perren}, Gabriel I. and {Pickering}, Timothy E. and {Rastogi}, Tanuj and {Roulston}, Benjamin R. and {Ryan}, Daniel F. and {Rykoff}, Eli S. and {Sabater}, Jose and {Sakurikar}, Parikshit and {Salgado}, Jes{'u}s and {Sanghi}, Aniket and {Saunders}, Nicholas and {Savchenko}, Volodymyr and {Schwardt}, Ludwig and {Seifert-Eckert}, Michael and {Shih}, Albert Y. and {Jain}, Anany Shrey and {Shukla}, Gyanendra and {Sick}, Jonathan and {Simpson}, Chris and {Singanamalla}, Sudheesh and {Singer}, Leo P. and {Singhal}, Jaladh and {Sinha}, Manodeep and {Sip{H{o}}cz}, Brigitta M. and {Spitler}, Lee R. and {Stansby}, David and {Streicher}, Ole and {{{S}}umak}, Jani and {Swinbank}, John D. and {Taranu}, Dan S. and {Tewary}, Nikita and {Tremblay}, Grant R. and {Val-Borro}, Miguel de and {Van Kooten}, Samuel J. and {Vasovi{'c}}, Zlatan and {Verma}, Shresth and {de Miranda Cardoso}, Jos{'e} Vin{'i}cius and {Williams}, Peter K.~G. and {Wilson}, Tom J. and {Winkel}, Benjamin and {Wood-Vasey}, W.~M. and {Xue}, Rui and {Yoachim}, Peter and {Zhang}, Chen and {Zonca}, Andrea and {Astropy Project Contributors}},
                title = "{The Astropy Project: Sustaining and Growing a Community-oriented Open-source Project and the Latest Major Release (v5.0) of the Core Package}",
              journal = {\apj},
                 year = 2022,
                month = aug,
               volume = {935},
               number = {2},
                  eid = {167},
                pages = {167},
                  doi = {10.3847/1538-4357/ac7c74},
        archivePrefix = {arXiv},
               eprint = {2206.14220},
         primaryClass = {astro-ph.IM},
               adsurl = {https://ui.adsabs.harvard.edu/abs/2022ApJ...935..167A}
        }

@article{abuterFirstLightGRAVITY2026,
  title = {First Light for the {{GRAVITY}}+ {{Adaptive Optics}}: {{Extreme}} Adaptive Optics for the {{Very Large Telescope Interferometer}}},
  shorttitle = {First Light for the {{GRAVITY}}+ {{Adaptive Optics}}},
  author = {Abuter, R. and Allouche, F. and Bailet, C. and Benisty, M. and Berdeu, A. and Berger, J.-P. and Berio, P. and Bigioli, A. and Blanchard, C. and Boebion, O. and Bonnet, H. and Bourdarot, G. and Bourget, P. and Brandner, W. and Brul{\'e}, J. and Burgos, P. and Carbillet, M. and Correia, C. and {Courtney-Barrer}, B. and Curaba, S. and Davies, R. and Defr{\`e}re, D. and Delboulb{\'e}, A. and Delplancke, F. and Dembet, R. and Drescher, A. and Dubost, N. and Eckart, A. and {\'E}douard, C. and Eisenhauer, F. and Otal, L. Esteras and Fabricius, M. and Feuchtgruber, H. and F{\'e}dou, P. and Finger, G. and Schreiber, N. M. F{\"o}rster and Frahm, R. and Garcia, E. and Garcia, P. and Lopez, R. Garcia and Genzel, R. and Gil, J. P. and Gillessen, S. and Gomes, T. and Gont{\'e}, F. and Gopinath, V. and Gouvret, C. and Graf, J. and Guajardo, P. and Guieu, S. and Hackenberg, W. and Hartl, M. and Haubois, X. and Hau{\ss}mann, F. and Henning, T. and Hibon, P. and H{\"o}nig, S. and Horrobin, M. and Houll{\'e}, M. and Hubin, N. and Taieb, I. Ibn and Jochum, L. and Jocou, L. and Jost, A. and Kammerer, J. and Karl, L. and Kaufer, A. and Kern, P. and Kervella, P. and Kolb, J. and Korhonen, H. and Kreidberg, L. and Krempl, P. and Lacour, S. and Lagarde, S. and Lai, O. and Lapeyr{\`e}re, V. and Laugier, R. and Leal, V. and Bouquin, J.-B. Le and Leftley, J. and L{\'e}na, P. and Lopez, B. and Lutz, D. and Magnard, Y. and Mang, F. and Marcotto, A. and Maurel, D. and M{\'e}rand, A. and Millour, F. and Montarges, M. and More, N. and Moruj{\~a}o, N. and Moulin, T. and Nowacki, H. and Nowak, M. and Oberti, S. and Ott, T. and Pallanca, L. and Patru, F. and Paumard, T. and Perraut, K. and Perrin, G. and Petrucci, P. O. and Petrov, R. and Pfuhl, O. and Pourr{\'e}, N. and Rabien, S. and Rau, C. and Riquelme, M. and {Robbe-Dubois}, S. and Rochat, S. and Salman, M. and {S{\'a}nchez-Berm{\'u}dez}, J. and Schubert, J. and Scigliuto, J. and Shchekaturov, P. and Schuhler, N. and Shangguan, J. and Shimizu, T. and Scheithauer, S. and Soenke, C. and Soulez, F. and Stadler, E. and Stadler, J. and Straubmeier, C. and Sturm, E. and Subroweit, M. and Sykes, C. and Tacconi, L. J. and Tristram, K. R. W. and Uysal, S. and von Fellenberg, S. and Widmann, F. and Wieprecht, E. and Wiezorrek, E. and Woillez, J. and Yazici, S. and Zins, G.},
  year = 2026,
  month = mar,
  journal = {Astronomy \& Astrophysics},
  volume = {707},
  pages = {A115},
  publisher = {EDP Sciences},
  issn = {0004-6361, 1432-0746},
  doi = {10.1051/0004-6361/202555666},
  urldate = {2026-08-23},
  copyright = {\copyright{} The Authors 2026},
  langid = {english}
}

@article{gotbergLongtermStabilityHR2016a,
  title = {Long-Term Stability of the {{HR}} 8799 Planetary System without Resonant Lock},
  author = {G{\"o}tberg, Ylva and Davies, Melvyn B. and Mustill, Alexander J. and Johansen, Anders and Church, Ross P.},
  year = 2016,
  month = aug,
  journal = {Astronomy \& Astrophysics},
  volume = {592},
  pages = {A147},
  publisher = {EDP Sciences},
  issn = {0004-6361, 1432-0746},
  doi = {10.1051/0004-6361/201526309},
  urldate = {2026-08-23},
  copyright = {\copyright{} ESO, 2016},
  langid = {english}
}

@article{brandtFirstDynamicalMass2021,
  title = {The {{First Dynamical Mass Measurement}} in the {{HR}} 8799 {{System}}},
  author = {Brandt, G. Mirek and Brandt, Timothy D. and Dupuy, Trent J. and Michalik, Daniel and Marleau, Gabriel-Dominique},
  year = 2021,
  month = jul,
  journal = {The Astrophysical Journal Letters},
  volume = {915},
  number = {1},
  pages = {L16},
  publisher = {The American Astronomical Society},
  issn = {2041-8205},
  doi = {10.3847/2041-8213/ac0540},
  urldate = {2026-08-23},
  langid = {english}
}

@article{maireLEECHExoplanetImaging2015,
  title = {The {{LEECH Exoplanet Imaging Survey}}. {{Further}} Constraints on the Planet Architecture of the {{HR}} 8799 System},
  author = {Maire, A.-L. and Skemer, A. J. and Hinz, P. M. and Desidera, S. and Esposito, S. and Gratton, R. and Marzari, F. and Skrutskie, M. F. and Biller, B. A. and Defr{\`e}re, D. and Bailey, V. P. and Leisenring, J. M. and Apai, D. and Bonnefoy, M. and Brandner, W. and Buenzli, E. and Claudi, R. U. and Close, L. M. and Crepp, J. R. and Rosa, R. J. De and Eisner, J. A. and Fortney, J. J. and Henning, T. and Hofmann, K.-H. and Kopytova, T. G. and Males, J. R. and Mesa, D. and Morzinski, K. M. and Oza, A. and Patience, J. and Pinna, E. and Rajan, A. and Schertl, D. and Schlieder, J. E. and Su, K. Y. L. and Vaz, A. and {Ward-Duong}, K. and Weigelt, G. and Woodward, C. E.},
  year = 2015,
  month = apr,
  journal = {Astronomy \& Astrophysics},
  volume = {576},
  pages = {A133},
  publisher = {EDP Sciences},
  issn = {0004-6361, 1432-0746},
  doi = {10.1051/0004-6361/201425185},
  urldate = {2026-08-23},
  copyright = {\copyright{} ESO, 2015},
  langid = {english}
}

@article{skemerFIRSTLIGHTLBT2012,
  title = {{{FIRST LIGHT LBT AO IMAGES OF HR}} 8799 Bcde {{AT}} 1.6 {{AND}} 3.3 {$\mu$}m: {{NEW DISCREPANCIES BETWEEN YOUNG PLANETS AND OLD BROWN DWARFS}}*},
  shorttitle = {{{FIRST LIGHT LBT AO IMAGES OF HR}} 8799 Bcde {{AT}} 1.6 {{AND}} 3.3 {$\mu$}m},
  author = {Skemer, Andrew J. and Hinz, Philip M. and Esposito, Simone and Burrows, Adam and Leisenring, Jarron and Skrutskie, Michael and Desidera, Silvano and Mesa, Dino and Arcidiacono, Carmelo and Mannucci, Filippo and Rodigas, Timothy J. and Close, Laird and McCarthy, Don and Kulesa, Craig and Agapito, Guido and Apai, Daniel and Argomedo, Javier and Bailey, Vanessa and Boutsia, Konstantina and Briguglio, Runa and Brusa, Guido and Busoni, Lorenzo and Claudi, Riccardo and Eisner, Joshua and Fini, Luca and Follette, Katherine B. and Garnavich, Peter and Gratton, Raffaele and Guerra, Juan Carlos and Hill, John M. and Hoffmann, William F. and Jones, Terry and Krejny, Megan and Males, Jared and Masciadri, Elena and Meyer, Michael R. and Miller, Douglas L. and Morzinski, Katie and Nelson, Matthew and Pinna, Enrico and Puglisi, Alfio and Quanz, Sascha P. and {Quiros-Pacheco}, Fernando and Riccardi, Armando and Stefanini, Paolo and Vaitheeswaran, Vidhya and Wilson, John C. and Xompero, Marco},
  year = 2012,
  month = jun,
  journal = {The Astrophysical Journal},
  volume = {753},
  number = {1},
  pages = {14},
  publisher = {The American Astronomical Society},
  issn = {0004-637X},
  doi = {10.1088/0004-637X/753/1/14},
  urldate = {2026-08-23},
  langid = {english}
}

@article{maroisDirectImagingMultiple2008c,
  title = {Direct {{Imaging}} of {{Multiple Planets Orbiting}} the {{Star HR}} 8799},
  author = {Marois, Christian and Macintosh, Bruce and Barman, Travis and Zuckerman, B. and Song, Inseok and Patience, Jennifer and Lafreni{\`e}re, David and Doyon, Ren{\'e}},
  year = 2008,
  month = nov,
  journal = {Science},
  volume = {322},
  number = {5906},
  pages = {1348--1352},
  publisher = {American Association for the Advancement of Science},
  doi = {10.1126/science.1166585},
  urldate = {2026-06-26}
}

@article{madhusudhanMODELATMOSPHERESMASSIVE2011,
  title = {{{MODEL ATMOSPHERES FOR MASSIVE GAS GIANTS WITH THICK CLOUDS}}: {{APPLICATION TO THE HR}} 8799 {{PLANETS AND PREDICTIONS FOR FUTURE DETECTIONS}}},
  shorttitle = {{{MODEL ATMOSPHERES FOR MASSIVE GAS GIANTS WITH THICK CLOUDS}}},
  author = {Madhusudhan, Nikku and Burrows, Adam and Currie, Thayne},
  year = 2011,
  month = jul,
  journal = {The Astrophysical Journal},
  volume = {737},
  number = {1},
  pages = {34},
  publisher = {The American Astronomical Society},
  issn = {0004-637X},
  doi = {10.1088/0004-637X/737/1/34},
  urldate = {2026-08-23},
  langid = {english}
}

@article{fabryckySTABILITYDIRECTLYIMAGED2010,
  title = {{{STABILITY OF THE DIRECTLY IMAGED MULTIPLANET SYSTEM HR}} 8799: {{RESONANCE AND MASSES}}},
  shorttitle = {{{STABILITY OF THE DIRECTLY IMAGED MULTIPLANET SYSTEM HR}} 8799},
  author = {Fabrycky, Daniel C. and {Murray-Clay}, Ruth A.},
  year = 2010,
  month = jan,
  journal = {The Astrophysical Journal},
  volume = {710},
  number = {2},
  pages = {1408},
  publisher = {The American Astronomical Society},
  issn = {0004-637X},
  doi = {10.1088/0004-637X/710/2/1408},
  urldate = {2026-08-23},
  langid = {english}
}

@ARTICLE{G23H,
       author = {{Thompson}, William and {Blakely}, Dori and {Xuan}, Jerry W. and {Blouin}, Simon and {Zhang}, Jingwen and {Johnstone}, Doug and {Ruffio}, Jean-Baptiste and {Nielsen}, Eric and {Speedie}, Jessica and {Bowler}, Brendan P. and {Bouchard-C{\^o}t{\'e}}, Alexandre and {Franson}, Kyle and {Blunt}, Sarah and {Roberson}, William and {Cloutier}, Ryan and {Fogal}, Andre and {Hessel}, Kaitlyn and {Marois}, Christian and {Rochon}, Alexandra},
        title = "{Detecting and Characterizing Companions with a Calibrated Gaia DR2, DR3, and Hipparcos Catalog (G23H)}",
      journal = {\aj},
         year = 2026,
        month = jul,
       volume = {172},
       number = {1},
          eid = {53},
        pages = {53},
          doi = {10.3847/1538-3881/ae64ed},
       adsurl = {https://ui.adsabs.harvard.edu/abs/2026AJ....172...53T}
}

@article{kieferSearchingSubstellarCompanion2025,
  title = {Searching for Substellar Companion Candidates with {{Gaia}} - {{I}}. {{Introducing}} the {{GaiaPMEX}} Tool},
  author = {Kiefer, F. and Lagrange, A.-M. and Rubini, P. and Philipot, F.},
  year = 2025,
  month = oct,
  journal = {Astronomy \& Astrophysics},
  volume = {702},
  pages = {A76},
  publisher = {EDP Sciences},
  issn = {0004-6361, 1432-0746},
  doi = {10.1051/0004-6361/202449335},
  urldate = {2026-08-28},
  copyright = {\copyright{} The Authors 2025},
  langid = {english}
}

@inproceedings{sivaramakrishnanPlanetarySystemStar2009a,
  title = {Planetary System and Star Formation Science with Non-Redundant Masking on {{JWST}}},
  booktitle = {Techniques and {{Instrumentation}} for {{Detection}} of {{Exoplanets IV}}},
  author = {Sivaramakrishnan, Anand and Tuthill, Peter G. and Ireland, Michael J. and Lloyd, James P. and Martinache, Frantz and Soummer, R{\'e}mi and Makidon, Russell B. and Doyon, Ren{\'e} and Beaulieu, Mathilde and Beichman, Charles A.},
  year = 2009,
  month = aug,
  volume = {7440},
  pages = {311},
  publisher = {SPIE},
  doi = {10.1117/12.826633},
  urldate = {2026-09-03},
  langid = {english}
}

@article{wangWhereistheplanetPredictingPositions2021,
  title = {Whereistheplanet: {{Predicting}} Positions of Directly Imaged Companions},
  shorttitle = {Whereistheplanet},
  author = {Wang, Jason J. and Kulikauskas, Matas and Blunt, Sarah},
  year = 2021,
  month = jan,
  journal = {Astrophysics Source Code Library},
  pages = {ascl:2101.003},
  urldate = {2026-09-07}
}
\bibliographystyle{aasjournalv7.1}

\end{document}